\documentclass[
 preprint,
superscriptaddress,
 amsmath,
 amssymb,
 amsthm,
  aps,
 prb,
floatfix,
natbib,
showpacs
]{revtex4-1}

\usepackage{graphicx,dcolumn,bm,gensymb,siunitx}

\begin{document}

\preprint{APS/123-QED}

\title{Structural Transformations of Li$_2$C$_2$ at High Pressures}

\author{Ilias Efthimiopoulos}
 \affiliation{Max-PIanck-Institute for Solid State Research, Heisenbergstrasse 1, D-70569 Stuttgart, Germany}
\author{Daryn E. Benson}
\affiliation{Department of Physics, Arizona State University, Tempe, Arizona 85287-1504}
\author{Sumit Konar}
 \affiliation{Department of Materials and Environmental Chemistry, Stockholm University, SE-10691 Stockholm, Sweden}
 \author{Johanna Nyl{\'e}n}
 \affiliation{Department of Materials and Environmental Chemistry, Stockholm University, SE-10691 Stockholm, Sweden}
 \author{Gunnar Svensson}
 \affiliation{Department of Materials and Environmental Chemistry, Stockholm University, SE-10691 Stockholm, Sweden}
\author{Ulrich H{\"a}ussermann}
 \affiliation{Department of Materials and Environmental Chemistry, Stockholm University, SE-10691 Stockholm, Sweden}
 \author{Stefan Liebig}
 \affiliation{Department of Chemistry, University of Cologne, Greinstrasse 6, 50939 Cologne, Germany}
 \author{Uwe Ruschewitz}
 \affiliation{Department of Chemistry, University of Cologne, Greinstrasse 6, 50939 Cologne, Germany}
 \author{Grigory V. Vazhenin}
 \affiliation{Max-PIanck-Institute for Solid State Research, Heisenbergstrasse 1, D-70569 Stuttgart, Germany}
 \author{Ingo Loa}
 \affiliation{The University of Edinburgh, School of Physics and Astronomy, Centre for Science at Extreme Conditions (CSEC), King's Buildings, Edinburgh EH9 3FD, United Kingdom}
 \author{Michael Hanfland}
 \affiliation{European Synchrotron Radiation Facility, 6 Rue Jules Horowitz, F 38000 Grenoble, France}
 \author{Karl Syassen}
 \affiliation{Max-PIanck-Institute for Solid State Research, Heisenbergstrasse 1, D-70569 Stuttgart, Germany}

\date{Aug 19, 2015}

\begin{abstract}
Structural changes of Li$_2$C$_2$ under pressure were studied by synchrotron x-ray diffraction in a diamond anvil cell under hydrostatic conditions and by using evolutionary search methodology for crystal structure prediction. We show that the high pressure polymorph of Li$_2$C$_2$, which forms from the \textit{Immm} ground state structure (\textit{Z} = 2) at around 15 GPa, adopts an orthorhombic \textit{Pnma} structure with \textit{Z} = 4. Acetylide C$_2$ dumbbells characteristic of \textit{Immm}-Li$_2$C$_2$ are retained in \textit{Pnma}-Li$_2$C$_2$. The structure of \textit{Pnma}-Li$_2$C$_2$ relates closely to the anticotunnite type structure. C$_2$ dumbbell units are coordinated by 9 Li atoms, as compared to 8 in the antifluorite structure of \textit{Immm}-Li$_2$C$_2$. First principles calculations predict a transition of \textit{Pnma}-Li$_2$C$_2$ at 32 GPa to a topologically identical phase with a higher \textit{Cmcm} symmetry. The coordination of C$_2$ dumbbell units by Li atoms is increased to 11. The structure of \textit{Cmcm}-Li$_2$C$_2$ relates closely to the Ni$_2$In type structure. It is calculated that \textit{Cmcm}-Li$_2$C$_2$ becomes metallic at pressures above 40 GPa. In experiments, however, \textit{Pnma}-Li$_2$C$_2$ is susceptible to irreversible amorphization.
\end{abstract}


\maketitle


\section{introduction}

Carbides of alkali and alkaline earth metals typically occur as salt-like acetylides which consist of C$_{\text{2}}^{\text{2-}}$ dumbbell anions isoelectronic to dinitrogen.\cite{ruschewitz_2003} Recent theoretical studies suggested that acetylide carbides should transform to modifications with polymeric carbon structures at moderate pressures (below 10 GPa).\cite{chen_2010, benson_2013,li_2013,wang_2014,li_2014,*li_2015} The predicted ``polycarbides" consist of carbon polyanions with chain, ribbon, or layer structures which are stabilized by electrostatic interactions with surrounding cations. Such polyanions occur typically in Zintl phases and are well known for e.g. silicon and germanium. For carbon they represent a hitherto unknown chemical and structural feature. Polycarbides display distinct electronic structures and are predicted to be superconductors.\cite{benson_2013, li_2013, wang_2014}
\par
Yet the computational predictions deviate notably from results of experimental high pressure studies. Hitherto investigated Li$_2$C$_2$, CaC$_2$ and BaC$_2$ have in common that acetylide C$_2$ dumbells are retained until irreversible amorphization occurs at pressures far higher than the calculated transition pressures for polymeric carbide formation.\cite{ilias_2012, ilias_2012b,*ilias_2010a, nylen_2012} The discrepancy has been attributed to kinetic hindrance.\cite{benson_2013} Prior to amorphization BaC$_2$ and Li$_2$C$_2$ undergo structural transformations at around 4 and 15 GPa, respectively, in room temperature experiments.\cite{ilias_2012, ilias_2012b,*ilias_2010a, nylen_2012}  These transformations correspond to a ``conventional" increase of coordination numbers with pressure, leading to denser packings of cations and dumbbells. In the ambient pressure structure of BaC$_2$ Ba$^{\text{2+}}$ and C$_{\text{2}}^{2-}$ ions are six-coordinated and arranged as in the NaCl structure. The rhombohedral high pressure modification relates to the CsCl structure with both types of ions attaining an eight-fold coordination.\cite{ilias_2012} For Li$_2$C$_2$ the structure of the high pressure form has not been conclusively characterized.\cite{ilias_2012b,*ilias_2010a, nylen_2012}\par  
Here we present the elucidation of the high pressure behavior of Li$_2$C$_2$ from combined synchrotron diffraction experiments and crystal structure prediction methodology. To prevent the generation of enthalpically more favorable polymeric carbides in the computations, a constrained evolutionary algorithm was employed that enforced retention of C$_2$ dumbbell units at high pressures.\cite{zhu_2012} We further show that if amorphization of Li$_2$C$_2$ were suppressed, a high pressure form predicted here would approach metallic behavior at pressures above 40 GPa. 

\section{methods}

\subsection{\label{sec:level2}Experiments}

All steps of sample preparation were performed in an Ar filled glove box  (H$_2$O and O$_2$ concentration $<$ 1 ppm). Starting materials for Li$_2$C$_2$ synthesis were lithium (ABCR, 99.99\%) and graphite powder (Sigma-Aldrich, 99.9998\%), which was degassed at 800 $\celsius$  under dynamic vacuum for 24 h prior to use. Stoichiometric amounts of lithium and graphite were transferred into a purified Ta ampoule. Afterwards the ampoule was sealed in He atmosphere (800 mbar) and was placed inside a quartz ampoule, which was sealed under vacuum. The quartz ampoule was heated for 24 hours at 1073 K in air (tube furnace) after which the sample was allowed to cool  by turning off the furnace. An air and moisture sensitive fine powder with a light-grey color was obtained. The phase purity of the sample was checked by  powder x-ray diffraction (PXRD, Huber G670, CuK$\alpha$1 radiation, capillary). Apart from a small amount of unreacted graphite, no impurities were detected.\par 
\textit{In situ} high pressure monochromatic PXRD experiments were performed with a membrane-driven diamond anvil cell (DAC) using a culet size of 400 microns. Powdered samples were loaded under inert gas atmosphere into a 150 micron-sized hole drilled in a stainless steel gasket. The pressure transmitting medium (PTM) was helium. Diffraction data were collected at room temperature at the ID09 beamline of the ESRF using a MAR555 flat panel detector. The x-ray wavelength was $\lambda$ = 0.41558 {\AA} and the beam diameter on the sample was set to  30 $\mu$m. In order to improve powder averaging, the DAC was rocked by $\pm3$ degrees. The pressure was monitored by the ruby luminescence method.\cite{syassen_2008} The two-dimensional diffraction data were integrated using the software Fit2D.\cite{hammersley_1996}\par 
All diffractograms were inspected using the STOE Win XPOW software system.\cite{winxpow_2010} DICVOL\cite{boultif_1991} within Win XPOW was used for indexing and ENDEAVOUR\cite{putz_1999} for an \textit{ab initio} structural solution using a direct space approach. Rietveld refinements were performed with GSAS.\cite{larson_2004} More details of the structure solution and refinement are given in Section III.

\subsection{Computations}

Structure searches were carried out using the evolutionary algorithm USPEX.\cite{oganov_2006,glass_2006, lyakhov_2013} The search over configurational space was constrained to structures containing C$_2$ acetylide units. C-C bond connectivity was enforced using the Z-matrix representation\cite{hoft_2006} available in the \textit{ab initio} code SIESTA.\cite{soler_2002} However, computationally demanding SIESTA was only used in the initial phase of a search as a means to quickly optimize the structure by constraining the molecular geometry and degrees of freedom of the C$_2$ acetylide units. These calculations employed the Perdew-Burke-Ernzerhof (PBE) exchange-correlation\cite{perdew_1996} as well as the single-$\zeta$ basis set. The plane wave cutoff was set at 100 Ry and a Monkhorst-Pack grid defined at a cutoff of 10 {\AA} was used. The pseudopotentials used were Troullier and Martins norm-conserving pseudopotentials.\cite{troullier_1991} The final stages of a search were performed using the Vienna Ab Initio Simulation Package (VASP).\cite{kresse_1996} The target pressure for searches was chosen to be 20 GPa. All populations contained 30 structures and the initial population's structures were randomly generated. All structures contained 16 atoms constrained to the chemical composition of Li$_2$C$_2$ (i.e. \textit{Z} = 4).\par
Enthalpy vs. pressure relations of Li$_2$C$_2$ phases were calculated using the first principles all-electron projector augmented waves (PAW) method\cite{blochl_1994a, *kresse_1999} as implemented in VASP.  Exchange-correlation effects were treated within the generalized gradient approximation (GGA) using the PBE parameterization.\cite{perdew_1996} The structures were relaxed with respect to pressure, lattice parameters, and atomic positions. Forces were converged to better than 1$\times$ $10^{-3}$ eV/{\AA}. The integration over the Brillouin Zone (BZ) was done on a grid of special k points of size 6 $\times$ 6 $\times$ 6, determined according to the Monkhorst-Pack scheme and using Gaussian smearing to determine the partial occupancies for each wavefunction.\cite{monkhorst_1976} The kinetic energy cutoff was set to 675 eV. To obtain the band structure and enthalpies the tetrahedron method with Bl\"{o}chl correction was employed for BZ integration.\cite{blochl_1994b} Structure relaxations and phonon calculations were performed at pressures ranging from 0 - 40 GPa. Once a structure was relaxed at a target pressure, zone-centered phonon calculations were executed using VASP's density functional perturbation theory approach.

\section{results and discussion}
\subsection{Experimental observations}

The ground state structure of Li$_2$C$_2$, \textit{Immm}-Li$_2$C$_2$, relates to the antifluorite structure. Li atoms are coordinated by four dumbbell units and each dumbbell unit by eight Li ions. When recording Raman spectra of Li$_2$C$_2$ in a DAC, it was consistently observed that \textit{Immm}-Li$_2$C$_2$ transforms reversibly at around 15 GPa into a high pressure modification. This is shown in Fig. 1. 

\begin{figure}
\includegraphics[width=\linewidth]{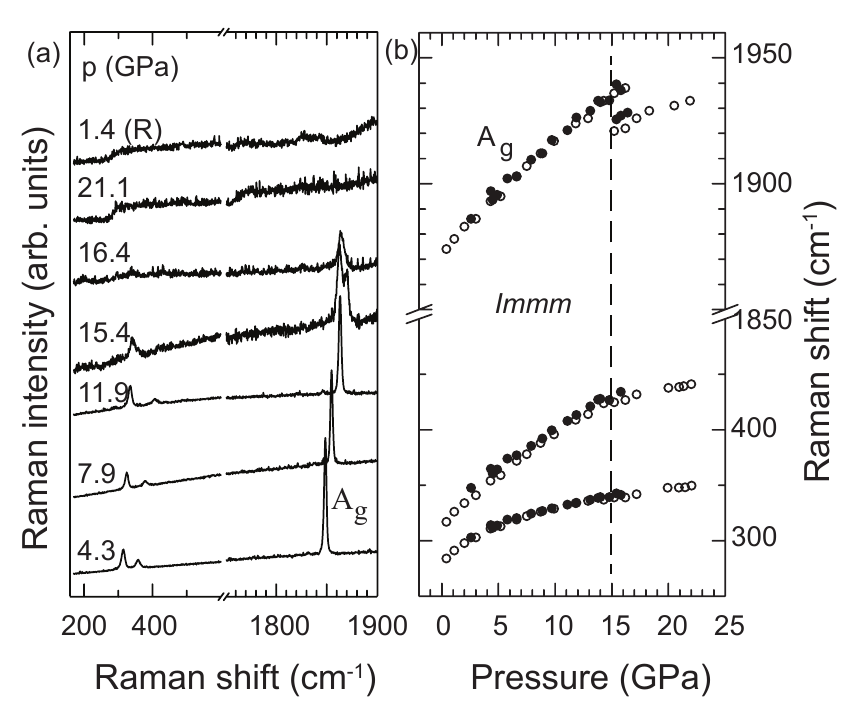}
 \caption{\label{fig1} (a) Raman spectra of polycrystalline Li$_2$C$_2$ at different pressures, (R) = decompression and (b) observed mode frequencies as a function of pressure from two experiments (black and white circles, respectively). The broken vertical line marks the transition pressure for a reversible structural transformation. No PTM was used in order to avoid scattering by sample surface contaminations. Li$_2$C$_2$ amorphizes irreversibly at pressures between 17 GPa (black circle experiment) and 24 GPa (white circle experiment, according to Ref. [9]).} 
 \end{figure}
 The retention of the dumbbell units is evidenced by the persistence of the acetylide C-C stretching vibration (A$_{\text{g}}$). The stretching mode frequency drops discontinuously by about 20 cm$^{\text{-1}}$ at the transition. At higher pressures Raman spectra became featureless, and remained featureless upon decompression.  This phenomenon is attributed to irreversible amorphization of Li$_2$C$_2$ at high pressures.\cite{nylen_2012}  In the Raman experiments no PTM was used in order to avoid any background scattering from possible surface contamination. The non-hydrostatic pressure conditions do not appear to influence the transition into the high pressure modification. However, pressures at which irreversible amorphization occurs varied between 17 and 25 GPa.\par

Figure 2 shows synchrotron PXRD patterns of Li$_2$C$_2$ across the phase transition and up to 24.7 GPa. 
\begin{figure}
\includegraphics[width=\linewidth]{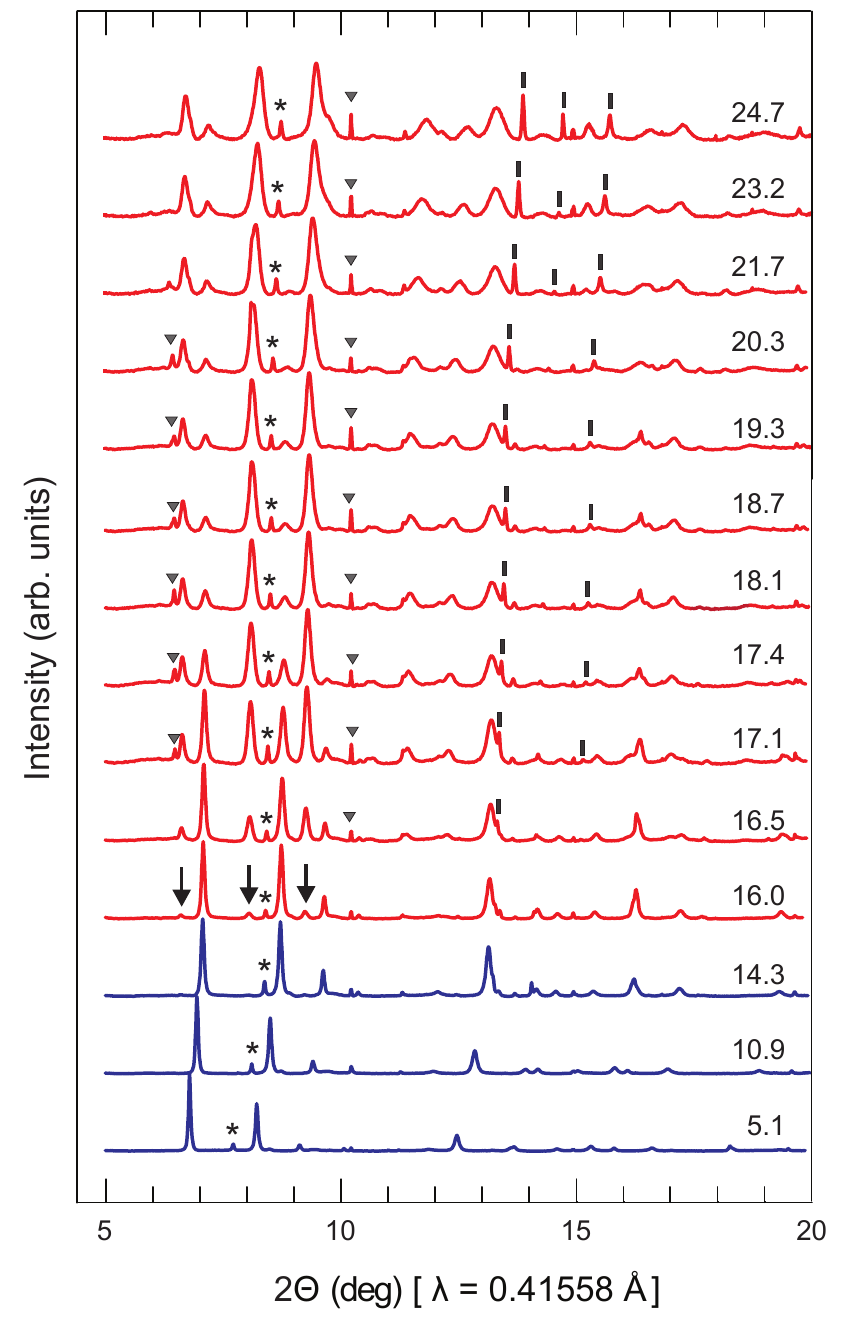}
 \caption{\label{fig2} (Color online) Compilation of x-ray diffraction patterns of Li$_2$C$_2$ ($\lambda$ = 0.41558 {\AA}) across the phase transition. Numbers are pressure in GPa. Blue patterns correspond to the pure \textit{Immm} phase. The arrows mark the appearance of \textit{Pnma} reflections. Asterisks mark a reflection from a graphite impurity. Triangles and bars mark reflections from ruby and the PTM He, respectively.} 
 \end{figure}
Different from the Raman studies, pressure conditions here were hydrostatic. Below 16 GPa patterns correspond to \textit{Immm}-Li$_2$C$_2$. At 16.5 GPa additional reflections appear. The onset pressure of the phase transition is in good agreement with the Raman experiments. \textit{Immm}-Li$_2$C$_2$ coexists with the high pressure modification as a non-equilibrium phase mixture beyond 20 GPa. The diffraction patterns taken at the highest applied pressure still indicated the presence of crystalline Li$_2$C$_2$, although reflections are broadened significantly. The data measured at 18.7 GPa were chosen for an \textit{ab initio} structure solution, as here the best resolution with respect to reflection overlap with \textit{Immm}-Li$_2$C$_2$ and broadening of reflections was found. The new diffraction peaks could be indexed with a primitive orthorhombic unit cell (a $\approx$ 5.1 {\AA}, b  $\approx$ 4.5 {\AA} , c $\approx$ 5.9 {\AA}), which pointed to \textit{Z} = 4. Due to the overlap of reflections a space group could not be determined unambiguously, but whole pattern decomposition suggested assignment of \textit{Pnma}. Using a direct space approach\cite{putz_1999} within this space group yielded a structural model that resembled the orthorhombic room temperature modification of Rb$_2$C$_2$ (\textit{Z} = 4).\cite{ruschewitz_2001}
\subsection{Elucidation of \textit{Pnma}-Li$_2$C$_2$}
To aid the structure elucidation, crystal structure searches by USPEX were performed at a target pressure of 20 GPa, well above the experimental transition pressure and below possible amorphization under hydrostatic conditions, respectively. Previous efforts using crystal structure prediction methodology in the structure search for high pressure Li$_2$C$_2$ have been restricted to simulation cells containing two formula units (i.e. 8 atoms).\cite{nylen_2012} This resulted in an energetically favorable structure (with \textit{Cmc}2$_{\text{1}}$ symmetry) for pressures above 15 GPa. However, calculated frequencies of Raman active modes for \textit{Cmc}2$_{\text{1}}$-Li$_2$C$_2$ deviated considerably from experiment. When extending the simulation cells to contain four formula units (16 atoms), as suggested by the diffraction experiments, the search yielded indeed a structure with \textit{Pnma} symmetry. Figure 3 shows the enthalpy differences (with respect to the \textit{Immm} ground state structure) as a function of pressure for \textit{Pnma}-Li$_2$C$_2$ and earlier predicted \textit{Cmc}2$_{\text{1}}$-Li$_2$C$_2$. 
\begin{figure}
\includegraphics[width=\linewidth]{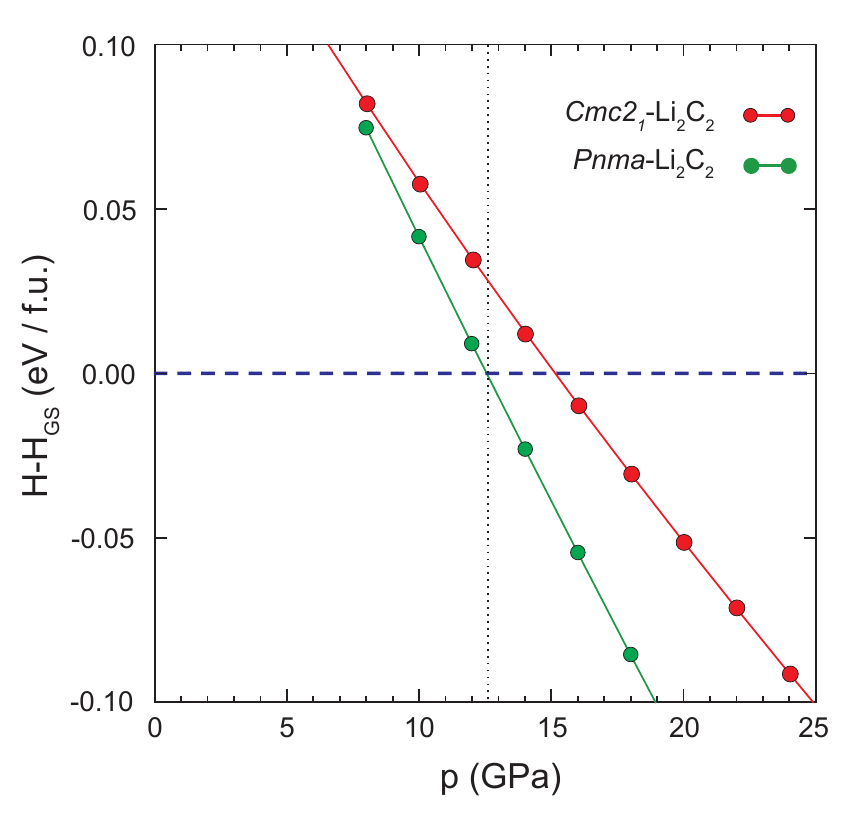}
 \caption{\label{fig3} (Color online) Calculated enthalpy-pressure relations (per formula unit) for Li$_2$C$_2$ with respect to the \textit{Immm} ground state structure. The dotted vertical line marks the transition pressure for the transformation to the \textit{Pnma} high pressure modification.}
 \end{figure}
 At pressures around 13 GPa the enthalpy of \textit{Pnma}-Li$_2$C$_2$ becomes lower than the ground state. This value for the transition pressure is slightly lower than the experimental observation. The minor discrepancy may be attributed to the negligence of zero-point-energy contributions and temperature effects in our calculations. Importantly, \textit{Pnma}-Li$_2$C$_2$ is dynamical stable in the pressure range 10 -- 30 GPa (see supplemental information\cite{notex}, Fig. S1). The structure parameters at 20 GPa are compiled in Tab. 1. 
 \begin{table}
\caption{\label{tab:table1}%
Structure parameters of \textit{Pnma}-Li$_2$C$_2$ at 20 GPa (DFT optimized) 
}
\begin{ruledtabular}
\begin{tabular}{llll}

Unit cell({\AA})	&	a = 5.0133	&	b = 4.4893	&	c = 5.8218	\\
Li1 (4c)	&	x = 0.1623	&	y = 0.25	&	z = 0.9033	\\
Li2 (4c)	&	x = 0.9945	&	y = 0.25	&	z = 0.2874	\\
C (8d) 	&	x = 0.7535	&	y = 0.1108	&	z = 0.9229	\\
\end{tabular}
\end{ruledtabular}
\end{table}
Additionally Ref. 29 contains parameters for the relaxed structures of \textit{Immm} and \textit{Pnma}-Li$_2$C$_2$  for the complete investigated pressure range 0--40 GPa (Tab. S1 and S2).\par 
For Rietveld refinement the structural parameters of the model obtained with USPEX were used as starting parameters. The refined parameters for \textit{Pnma}-Li$_2$C$_2$ at 18.7 GPa are given in Tab. 2.
\begin{table}
\caption{\label{tab:table2}%
Structure parameters of \textit{Pnma}-Li$_2$C$_2$ at 18.7 GPa (Rietveld refinement) 
}
\begin{ruledtabular}
\begin{tabular}{llll}

Unit cell({\AA})	&	a = 5.098(2)	&	b = 4.505(1)	&	c = 5.909(2)	\\

Li1 (4c)	&	x = 0.144(4)	&	y = 0.25	&	z = 0.938(3)	\\
Li2 (4c)	&	x = 0.999(5)	&	y = 0.25	&	z = 0.227(5)	\\
C (8d) 	&	x = 0.742(1)	&	y = 0.1163(3)	&	z = 0.9100(6)	\\
\end{tabular}
\end{ruledtabular}
\end{table}
 Details of the measurement and the refinement are summarized in Tab. S3.\cite{notex} In Tab. 3 interatomic distances are compared with those of the computed structure at 20 GPa.
\begin{table}
\caption{\label{tab:table3}%
Interatomic distances ({\AA}) in \textit{Pnma}-Li$_2$C$_2$
}
\begin{ruledtabular}
\begin{tabular}{ccc}

Atom pairs	&	Exp. structure 	&	Comp. structure 	\\
	&	(18.7 GPa) 	&	(20 GPa)	\\
	\hline
Li1 -- Li	&	1.86 -- 2.68 {\AA} (4$\times$)	&	2.39 -- 2.62 {\AA} (4$\times$)	\\
Li2 -- Li	&	1.86 -- 2.68 {\AA} (6$\times$)	&	2.39 -- 2.62 {\AA} (6$\times$)	\\
Li1 -- C	&	1.97 {\AA} (2$\times$), 	&	1.96 {\AA} (2$\times$), 	\\
	&	2.14 {\AA} (2$\times$), 	&	2.05 {\AA} (2×), 	\\
	&	2.20 {\AA} (2×)	&	2.15 {\AA} (2$\times$)	\\
Li2 -- C	&	2.26 {\AA} (2$\times$), 	&	2.19 {\AA} (2$\times$), 	\\
	&	2.33 {\AA} (2$\times$), 	&	2.22 {\AA} (2$\times$), 	\\
	&	2.36 {\AA} (2$\times$), 	&	2.39 {\AA} (2$\times$), 	\\
	&	2.55 {\AA} (2$\times$)	&	2.52 {\AA} (2$\times$)	\\
C -- C 	&	1.20\footnote{soft constraints} 	&	1.25	\\
C -- Li 	&	1.97 -- 2.55 (7$\times$)	&	1.96 -- 2.52 (7$\times$)	\\

\end{tabular}
\end{ruledtabular}
\end{table}
 Especially the refinement of the Li atom positions was quite unstable and lead to a few short Li-Li distances. This can be attributed to the modest data quality and the strong overlap of reflections from coexisting \textit{Immm}-Li$_2$C$_2$. The reduced data quality could be a consequence of Li disorder, connected to the occurrence of an intermediate phase between \textit{Immm}- and \textit{Pnma}-Li$_2$C$_2$. Such an intermediate phase has been identified for the high pressure phase transition of Li$_2$S\cite{grzechnik_2000} which, as we will discuss later, relates closely to that of Li$_2$C$_2$. Also, an intermediate phase with varying Li disorder might explain the extended range of coexistence of \textit{Immm}- and \textit{Pnma}-Li$_2$C$_2$. However, such a phase could not be unambiguously identified from our diffraction data. The final fit of the 18.7 GPa data is shown in Fig. 4. 
\begin{figure}
\includegraphics[width=\linewidth]{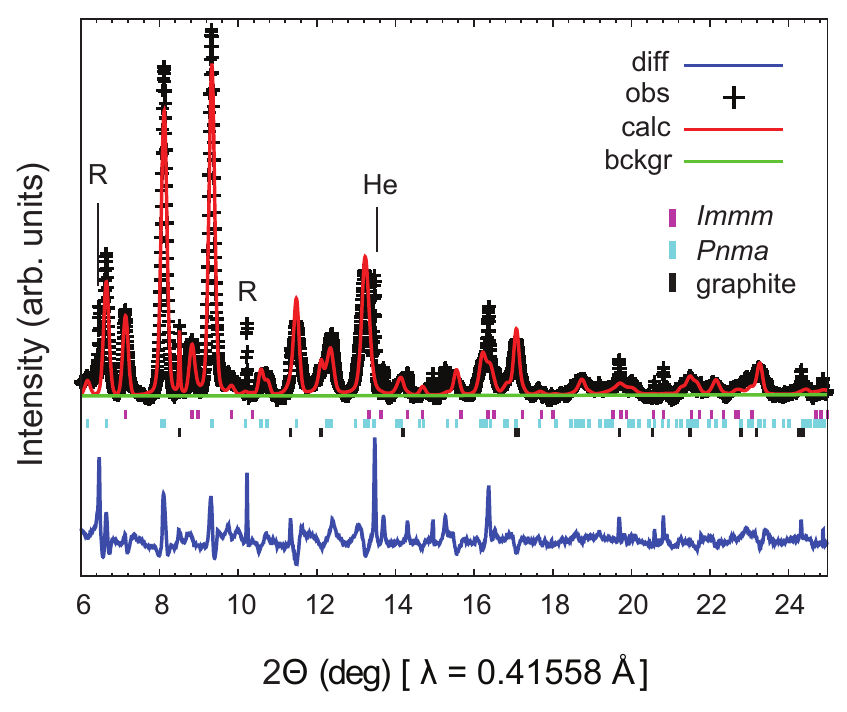}
 \caption{\label{fig4} (Color online) Rietveld refinement of the synchrotron PXRD pattern of \textit{Pnma}-Li$_2$C$_2$ at 18.7 GPa ($\lambda$ = 0.41558 {\AA}). Experimental data points (+), calculated profile (red solid line), and difference curve (blue curve below) are shown. Vertical bars mark the positions of Bragg reflections of graphite (black), \textit{Pnma}- (light blue) and \textit{Immm}-Li$_2$C$_2$ (magenta). Sharp extra reflections stem from ruby (R) and the PTM He.}
 \end{figure}
Differences between the calculated and measured profiles (in particular, extra sharp reflections) can mainly be attributed to ruby and solid helium. Attempts to improve the fit by applying parameters for stress, strain or anisotropic peak broadening gave unstable refinements and did not lead to physically meaningful results. Only the refinement of preferred orientation parameters (March-Dollase) gave a significant improvement of the fit. In Tab. S4 we also present the results from Rietveld refinements of the data at 7.2, 18.1, 18.7 and 19.3 GPa,respectively.\cite{notex}\par 
Figure 5 shows the pressure-volume (p-V) relations of \textit{Immm} and \textit{Pnma}-Li$_2$C$_2$.
\begin{figure}
\includegraphics[width=\linewidth]{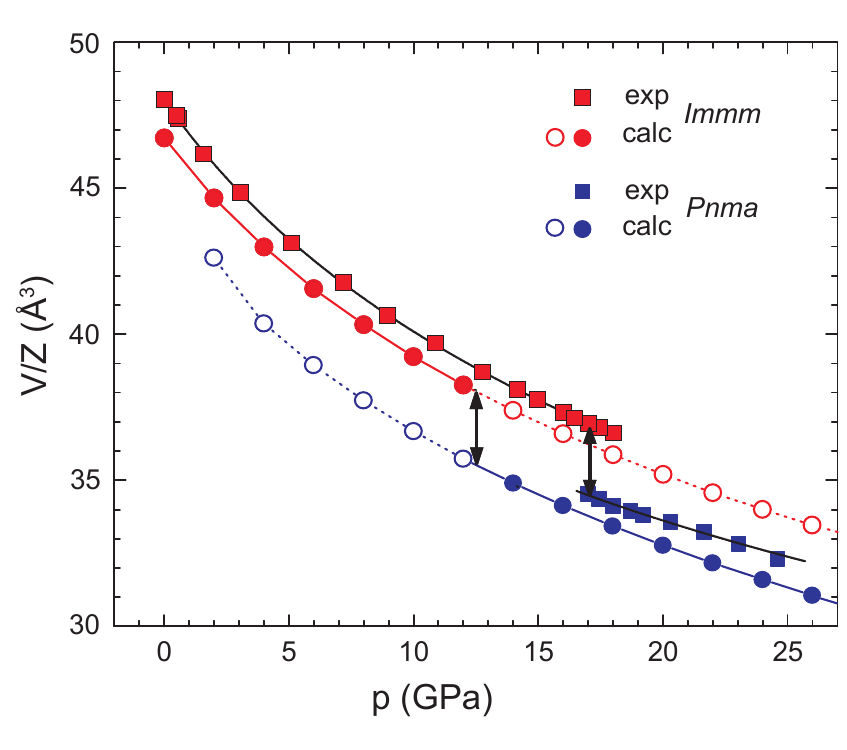}
 \caption{\label{fig5} (Color online) Volume versus pressure data for \textit{Immm} (red symbols) and \textit{Pnma}-Li$_2$C$_2$ (blue symbols). Experimental and computed values are presented as squares and circles, respectively. The transition pressures are marked by arrows.}
 \end{figure}
Unit cell parameters as a function of pressure from diffraction data are given in Tab. S5 and S6.\cite{notex} Both experimental and calculated p-V data were fitted to a three-parameter Birch-Murnaghan equation of state (EOS) expression.\cite{birch_1947} Generally there is good agreement between calculated and experimentally determined p-V data. Computed volumes are somewhat underestimated, by 2 -- 2.5 \%. The first order phase transition from \textit{Immm}- to \textit{Pnma}-Li$_2$C$_2$ is accompanied by a 7 \% volume reduction. The fitted EOS parameters are presented in Tab. 4.
\begin{table}
\caption{\label{tab:table4}%
Equation of state parameters for phases of Li$_2$C$_2$. Note that the experimental results for \textit{Pnma}-Li$_2$C$_2$ refer to a reference pressure of 16.5 GPa, not zero pressure.  
}
\begin{ruledtabular}
\begin{tabular}{cccc}

Li$_2$C$_2$	&	$V_0$ ({\AA}$^{\text{3}}$)	&	$K_0$ (GPa)	&	$K_0$'	\\
\hline
\textit{Immm} exp	&	47.9	&	39(1)	&	3.9(2)	\\
\textit{Pnma} exp	&	$V_r$ = 34.5	&	$K_r$ = 112(5)	&	4 (fixed)	\\
							
\textit{Immm} calc	&	46.7	&	40.8	&	3.9	\\
\textit{Pnma} calc	&	44.13	&	34.7	&	4.9	\\
\textit{Cmcm} calc	&	42.71	&	38.7	&	4.3	\\

\end{tabular}
\end{ruledtabular}
\end{table}
For \textit{Immm}-Li$_2$C$_2$ computed and experimental p-V data give virtually identical parameters. The ambient-pressure bulk modulus of this phase is around 40 GPa. For \textit{Pnma}-Li$_2$C$_2$ the bulk modulus extracted from the experimental data is 112 GPa at the reference pressure $p_r$ = 16.5 GPa ($V_r$ = 34.5 {\AA}$^{\text{3}}$). 

\subsection{\textit{Cmcm}-Li$_2$C$_2$ and structural relationships}
The high-pressure phase \textit{Pnma}-Li$_2$C$_2$ amorphizes irreversibly in room temperature Raman experiments at $\sim$17 GPa (non-hydrostatic) but persists up to at least 25 GPa under hydrostatic conditions. Computationally \textit{Pnma}-Li$_2$C$_2$ may be further compressed. Interestingly, as shown in Fig. 6, this yields at around 32 GPa a transition into another structure.
 \begin{figure}
 \includegraphics[width=\linewidth]{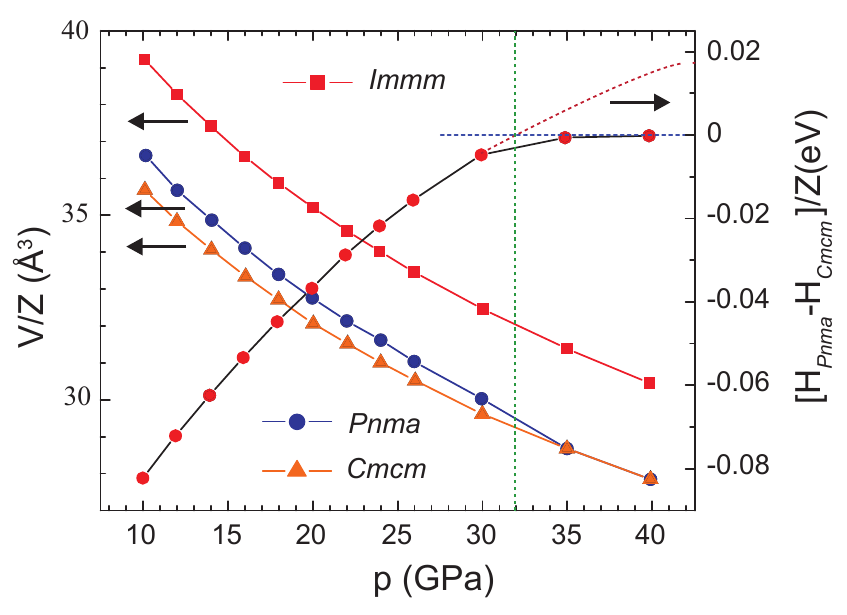}
 \caption{\label{fig6} (Color online) Volume-pressure relation (left ordinate) of the high pressure phases \textit{Pnma}-Li$_2$C$_2$ and \textit{Cmcm}-Li$_2$C$_2$ (the \textit{Immm} ground state structure is included for comparison) and enthalpy-pressure relation of \textit{Pnma}-Li$_2$C$_2$ with respect to \textit{Cmcm}-Li$_2$C$_2$ (right ordinate). The dotted red line is a polynomial fit of H$_{\textit{Pnma}}$-H$_{\textit{Cmcm}}$ to pressures $<$ 30 GPa. The transition pressure is marked by a vertical line.}
 \end{figure}
The new structure is topologically equivalent to \textit{Pnma}-Li$_2$C$_2$, but adopts the higher symmetry space group \textit{Cmcm}. The structure parameters for \textit{Cmcm}-Li$_2$C$_2$ and their variation with pressure are compiled in Tab. S7 in Ref. 29, EOS parameters are included in Table 4.\par 
Figure 7 depicts the structural relations between ground state \textit{Immm}-Li$_2$C$_2$ and the \textit{Pnma} and \textit{Cmcm} high pressure phases.
\begin{figure*}
 \includegraphics[width=\linewidth]{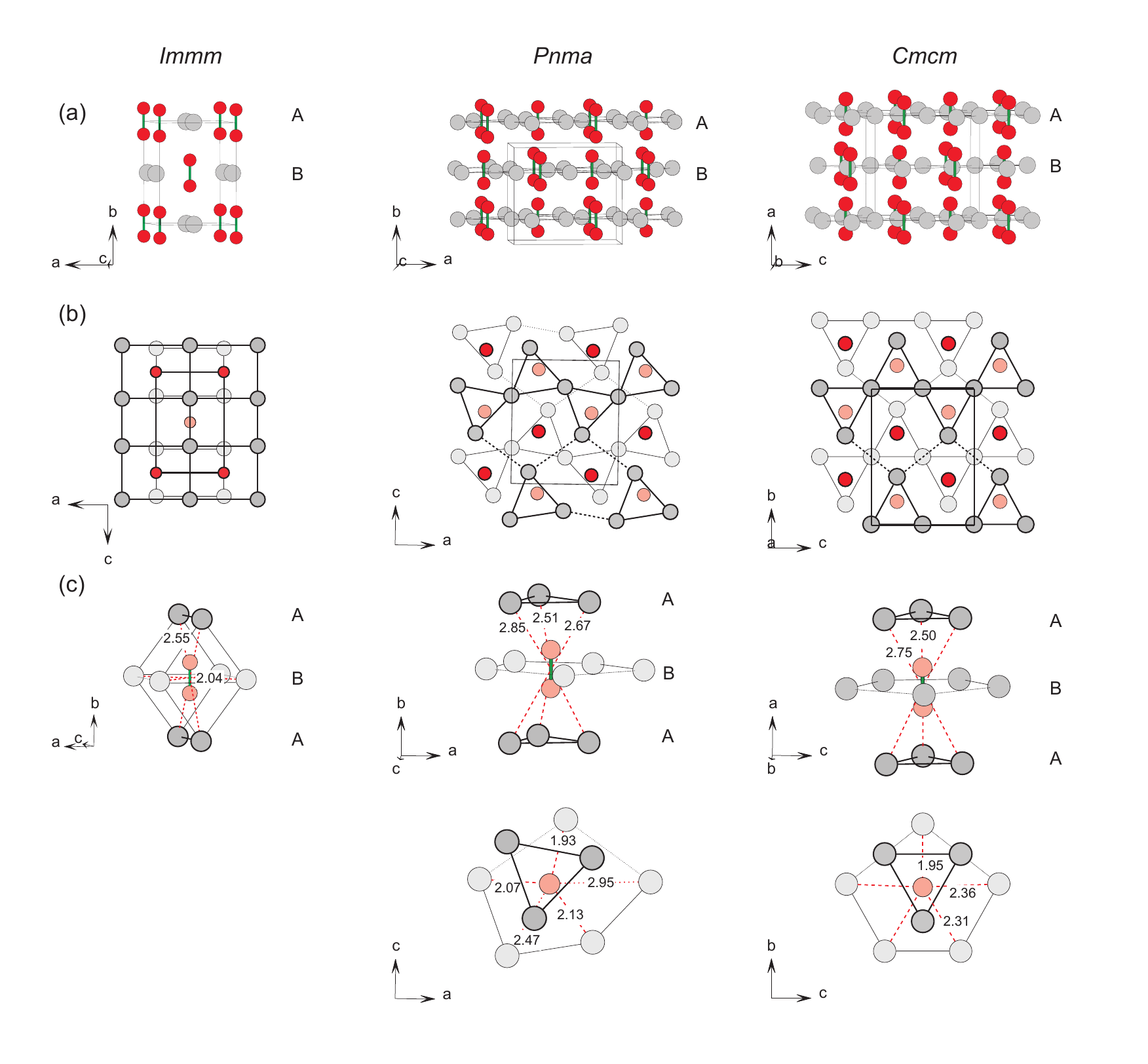}
\caption{\label{fig7}(Color online) (a) Crystal structures of \textit{Immm}, \textit{Pnma}, and \textit{Cmcm}-Li$_2$C$_2$ represented as layers consisting of planar nets formed by Li ions, which are centered by perpendicularly oriented dumbbell units. Li ions are shown as light grey circles and C atoms as red circles. Layers are stacked with an AB sequence in the dumbbell direction, as described in the text. (b) View of the structures along the layer stacking direction. A-type layers are distinguished by bold lines. B-Type layers by thin lines and pale colors. (c) Coordination of C$_2$ dumbbells within the three phases. The numbers indicate the distances between the dumbbell center and surrounding Li ions in {\AA} (referring to DFT optimized structures at 20 GPa).}
 \end{figure*}
 As mentioned earlier, the \textit{Immm} structure relates to the antifluorite type: C$_2$ dumbbells are arranged as a quasi cubic close packing in which Li atoms occupy the tetrahedral voids. Consequently, each C$_2$ dumbbell is surrounded by 8 Li atoms and each Li atom by 4 dumbbell units. Alternatively, the \textit{Immm} structure can be viewed as a stacking of layers consisting of planar, rectangular nets formed by the Li ions, which are stuffed by C$_2$ dumbbells oriented perpendicularly. Layers are stacked along the \textit{b} direction (which is the elongation direction of dumbbells) and consecutive layers A and B are related by the \textit{I} centering.\par
Also within \textit{Pnma}-Li$_2$C$_2$ Li ions form planar nets (parallel to the \textit{ac} plane), which consist of triangle ribbons running along the \textit{a} direction. Interatomic distances within triangles are short compared to distances in-between (2.5 {\AA} vs 3.1 {\AA} at 20 GPa). Planar Li nets are completed to layers by perpendicularly oriented C$_2$ dumbbells interspersed between triangle ribbons. In the \textit{Pnma} structure consecutive layers A and B are stacked in a way that C$_2$ dumbbells (e.g. in a layer A) attain a trigonal prismatic coordination by two triangles from adjacent layers above and below (layers B). A dumbbell is coordinated additionally by three Li ions which are situated in the same layer and cap the rectangular faces of the trigonal prism. Consequently, compared to \textit{Immm}-Li$_2$C$_2$ the coordination of a dumbbell by Li ions is increased to 9.\par 
As \textit{Immm}-Li$_2$C$_2$ relates to the antifluorite type so does the \textit{Pnma} structure to the anticotunnite type. \textit{Pnma}-Li$_2$C$_2$ is isostructural to the recently discovered ternary carbides CsKC$_2$ and CsRbC$_2$\cite{liebig_2013} and antifluorite -- anticotunnite transitions are frequently observed for alkali metal chalcogenides A$_2$B at high pressures. In particular, Li$_2$O and Li$_2$S display this transition at around 45 and 12 GPa, respectively.\cite{grzechnik_2000, kunc_2005} For Na$_2$S the antifluorite ground state structure transforms to the anticotunnite structure at even lower pressures, at around 7 GPa. At about 16 GPa another transition takes place which results in a phase with the Ni$_2$In type structure.\cite{vegas_2001,* vegas_2002, *santamaria_2011}\par 
Interestingly, the sequence antifluorite $\rightarrow$ anticontunnite $\rightarrow$ Ni$_2$In type is also shown by Li$_2$C$_2$ as \textit{Cmcm}-Li$_2$C$_2$ relates to the hexagonal Ni$_2$In structure. The topology of planar Li ion nets is identical in \textit{Pnma} and \textit{Cmcm}-Li$_2$C$_2$. However, in the higher symmetry \textit{Cmcm} structure ribbons are straightened into distinct zigzag chains in which triangles are strictly oriented up and down (cf. Fig. 7(b)). These chains run along the \textit{c} direction. The orientation of triangles from adjacent chains yields five-membered rings, which are centered by the dumbbell units. Because the trigonal prismatic environment of a dumbbell by Li triangles situated in layers above and below is maintained, its total coordination by Li ions is increased to 11 with respect to the \textit{Pnma} structure. The coordination polyhedron corresponds to an Edshammar polyhedron which is the signature of the Ni$_2$In structure type.\cite{hyde_1989}\par
To conclude the discussion of structural relationships, we address the evolution of interatomic distances with pressure (referring to the DFT optimized structures). The C-C distance within dumbbell units is only slightly compressible. Within the \textit{Immm} structure this distance reduces from 1.256 {\AA} at ambient pressure to 1.239 {\AA} at 40 GPa. This is similar for the high pressure forms. Here this distance decreases from 1.254 {\AA} at 10 GPa to 1.244 {\AA} at 40 GPa. The Li-Li distances defining the coordination around C$_2$ dumbbells are 2.55, 2.81 and 3.02 {\AA} for the \textit{Immm} structure at ambient pressure. They reduce to 2.35, 2.58, and 2.84 {\AA} at 14 GPa which is close to the calculated transition pressure. At this pressure the corresponding Li-Li distances in the \textit{Pnma} structure are between 2.46 and 3.17 {\AA}.

\subsection{Electronic structure changes with pressure}

The band structures of \textit{Immm}-Li$_2$C$_2$ and \textit{Pnma}/\textit{Cmcm}-Li$_2$C$_2$ are shown in Fig. 8.
\begin{figure}
\includegraphics[width=\linewidth]{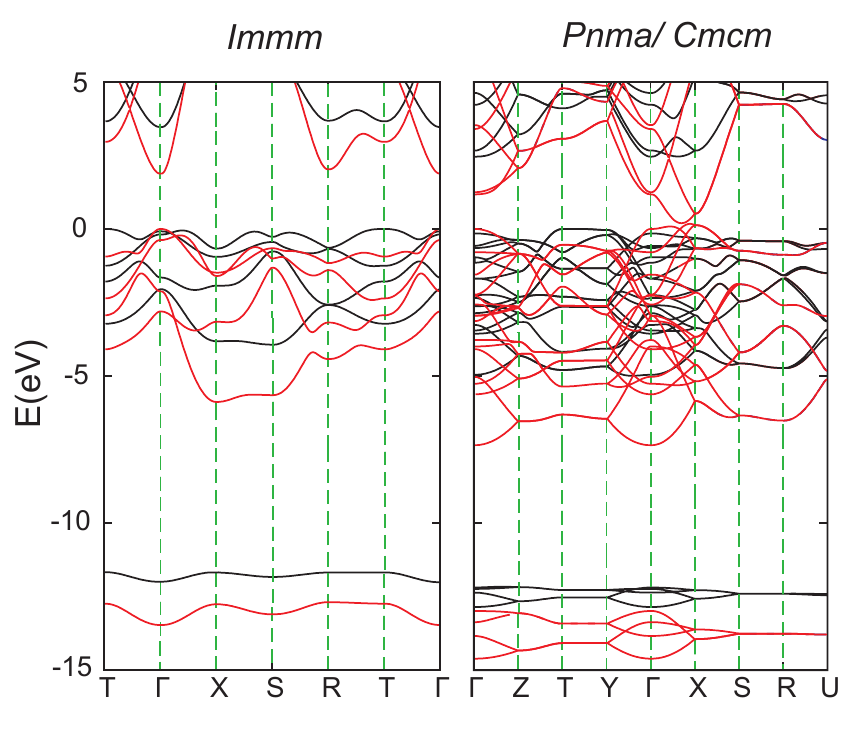}
 \caption{\label{fig8}(Color online) Calculated band structures of the ground state (left) and the higher pressure phases (right) of Li$_2$C$_2$. Black lines represent the ground state and \textit{Pnma} high pressure phase at zero and eight GPa, respectively. Red lines represent the ground state and \textit{Cmcm} high pressure phase at 40 GPa.}
 \end{figure}
At pressures below 10 GPa both the ambient- and high-pressure forms exhibit insulating properties. At ambient pressure \textit{Immm}-Li$_2$C$_2$ has an indirect band gap of 3.3 eV with the bottom of the conduction band at $\Gamma$ and the top of the valence band at T. The valence bands mirror the molecular orbital (MO) diagram of the acetylide anion. Their topology for Li$_2$C$_2$ is similar to CaC$_2$ whose electronic structure has been studied earlier.\cite{ruiz_1995, long_1992} The weakly dispersed band centered at -12 eV below the Fermi level corresponds to the sp$\sigma_{\text{g}}$ bonding MO. Bands corresponding to the two lone pair states (sp$\sigma_{\text{u}}$ and sp$\sigma_{\text{g}}$) have dispersions of about 2 eV and are located in the range -4 to -1 eV below the Fermi level. The two $\pi$-bonding bands constitute the top of the valence band. It is clearly seen that pressure increases especially the lone-pair -- Li interactions because the dispersion of lone-pair based bands increases most. The pressure dependence of the DFT-GGA computed band gap is shown in Fig. 9.
 \begin{figure}
  \includegraphics[width=\linewidth]{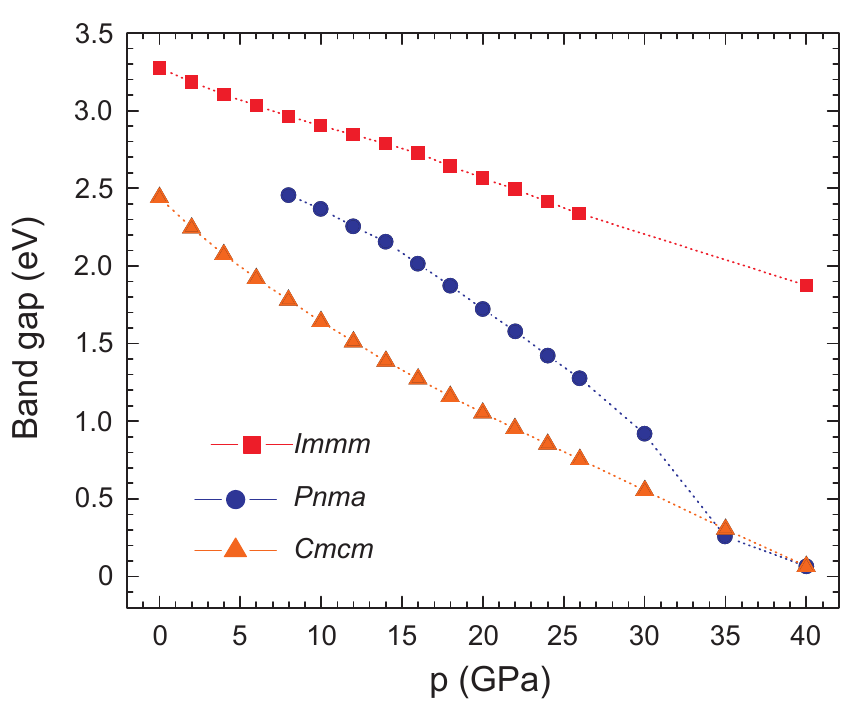}
 \caption{\label{fig9}(Color online) Band gap-pressure relations of the ground state and high pressure phases of Li$_2$C$_2$.} 
 \end{figure}
 It decreases linearly, but \textit{Immm}-Li$_2$C$_2$ obviously stays insulating.\par
At low pressure (below 10 GPa) \textit{Pnma}-Li$_2$C$_2$ exhibits an indirect band gap of $<$ 2.5 eV with the bottom of the conduction band at $\Gamma$ and the top of the valence band lying along T-Y. The band gap of \textit{Pnma}-Li$_2$C$_2$ diminishes faster with pressure compared to the \textit{Immm} structure. In high pressure Raman experiments a darkening of the sample is observed after the \textit{Immm} to \textit{Pnma} phase transition.\cite{nylen_2012} This possibly relates to the considerably decreased band gap of \textit{Pnma}-Li$_2$C$_2$. At 35 GPa the \textit{Pnma} structure merged into the \textit{Cmcm} one. At this pressure the calculated band gap dropped below 0.5 eV. Above 40 GPa the band gap of \textit{Cmcm}-Li$_2$C$_2$ has closed. The comparatively low pressure for (hypothetical) metallization of an ionic structure is remarkable. The changed pressure dependence of the band gap for the high pressure phases (compared to \textit{Immm}-Li$_2$C$_2$) can be attributed to the different coordination of dumbbell units. In the high pressure phases acetylide lone pairs are coordinated by triangles of Li ions. With pressure this coordination will develop into a covalent bonding interaction between C and Li, formally corresponding to electron donation from the dumbbell C$_{\text{2}}^{2-}$ to Li$^{+}$, and eventually leading to metallization.

\section{conclusions}

In summary, we have employed a combination of synchrotron x-ray diffraction experiments and computational evolutionary search methodology to elucidate the high pressure crystal structure of the acetylide carbide Li$_2$C$_2$. The observed high-pressure phase has \textit{Pnma} symmetry and relates to the anticotunnite structure (\textit{Z} = 4). In hydrostatic experiments \textit{Pnma}-Li$_2$C$_2$ does not amorphize under pressures up to 25 GPa. We find that if \textit{Pnma}-Li$_2$C$_2$ were prevented from amorphization it would transform at around 32 GPa to a higher symmetry \textit{Cmcm} structure which is closely related to the Ni$_2$In type. \textit{Cmcm}-Li$_2$C$_2$ would metalize at pressures above 40 GPa as a result of indirect band overlap. The sequence antifluorite$\rightarrow$ anticotunnite $\rightarrow$ Ni$_2$In type mirrors a common trend of high pressure phase transitions in A$_2$X compounds toward higher coordination.\par 
We point out that the high pressure behavior of the acetylide carbides Li$_2$C$_2$ and \textit{M}C$_2$ (M = Ca, Sr, Ba) appears strikingly similar to the corresponding sulfides. Experimental and/or calculated transition pressures for the sequences antifluorite $\rightarrow$ anticontunnite$\rightarrow$ Ni$_2$In type (referring to Li$_2$C$_2$/Li$_2$S) and rocksalt $\rightarrow$ CsCl type (referring to \textit{M}C$_2$/\textit{M}S) are remarkably close.\cite{ilias_2012, ilias_2012b, nylen_2012, schon_2004, yamaoka_1980, lu_2014, kulkarni_2010, syassen_1985, luo_1994} This may be attributed to a similar polarizability of the C$_{\text{2}}^{2-}$ and S$^{\text{2-}}$ anions. However, unlike sulfides, acetylides will undergo amorphization with pressure and expected phase transitions may not be observed. The origin of the pressure induced amorphization of acetylides is uncertain, and different scenarios can be envisioned. For example, amorphization could indicate compositional instability and phase segregation, which appears to be the case for BaC$_2$.\cite{wang_2014} Further, it could connect with a pressure limit for stability of multiple bonded light element entities, as suggested in Ref. 7. However, the enthalpic pressure limit for the stability of C$_{\text{2}}^{2-}$ units is rather low, as computations show clearly that with pressure carbides with polymeric anions become rapidly favored over acetylides. Specifically, for Li$_2$C$_2$ a phase with the CrB structure becomes more stable than \textit{Immm}-Li$_2$C$_2$ at already 5 GPa.\cite{chen_2010, benson_2013} This is far below the experimentally observed \textit{Immm} to \textit{Pnma} phase transition (see also Fig. S2 in Ref. 29) and it has been concluded that acteylides are distinguished by a high kinetic stability.\citep{benson_2013} The elucidation of the origin of the kinetic stability and pressure induced amorphization of acetylides will require the analysis of the composition and local structure of the amorphous carbides by e.g. synchrotron EXAFS and/or total scattering experiments, preferably in combination with molecular dynamics simulations.

\begin{acknowledgments}
This work was supported by the Swedish Research Council through grant 2012-2956 and the National Science Foundation through grant DMR-1007557.
\end{acknowledgments}

\bibliography{PRB-Li2C2}

\begin{thebibliography}{48}%
\makeatletter
\providecommand \@ifxundefined [1]{%
 \@ifx{#1\undefined}
}%
\providecommand \@ifnum [1]{%
 \ifnum #1\expandafter \@firstoftwo
 \else \expandafter \@secondoftwo
 \fi
}%
\providecommand \@ifx [1]{%
 \ifx #1\expandafter \@firstoftwo
 \else \expandafter \@secondoftwo
 \fi
}%
\providecommand \natexlab [1]{#1}%
\providecommand \enquote  [1]{``#1''}%
\providecommand \bibnamefont  [1]{#1}%
\providecommand \bibfnamefont [1]{#1}%
\providecommand \citenamefont [1]{#1}%
\providecommand \href@noop [0]{\@secondoftwo}%
\providecommand \href [0]{\begingroup \@sanitize@url \@href}%
\providecommand \@href[1]{\@@startlink{#1}\@@href}%
\providecommand \@@href[1]{\endgroup#1\@@endlink}%
\providecommand \@sanitize@url [0]{\catcode `\\12\catcode `\$12\catcode
  `\&12\catcode `\#12\catcode `\^12\catcode `\_12\catcode `\%12\relax}%
\providecommand \@@startlink[1]{}%
\providecommand \@@endlink[0]{}%
\providecommand \url  [0]{\begingroup\@sanitize@url \@url }%
\providecommand \@url [1]{\endgroup\@href {#1}{\urlprefix }}%
\providecommand \urlprefix  [0]{URL }%
\providecommand \Eprint [0]{\href }%
\providecommand \doibase [0]{http://dx.doi.org/}%
\providecommand \selectlanguage [0]{\@gobble}%
\providecommand \bibinfo  [0]{\@secondoftwo}%
\providecommand \bibfield  [0]{\@secondoftwo}%
\providecommand \translation [1]{[#1]}%
\providecommand \BibitemOpen [0]{}%
\providecommand \bibitemStop [0]{}%
\providecommand \bibitemNoStop [0]{.\EOS\space}%
\providecommand \EOS [0]{\spacefactor3000\relax}%
\providecommand \BibitemShut  [1]{\csname bibitem#1\endcsname}%
\let\auto@bib@innerbib\@empty
\bibitem [{\citenamefont {Ruschewitz}(2003)}]{ruschewitz_2003}%
  \BibitemOpen
  \bibfield  {author} {\bibinfo {author} {\bibfnamefont {U.}~\bibnamefont
  {Ruschewitz}},\ }\href@noop {} {\bibfield  {journal} {\bibinfo  {journal}
  {Coord. Chem. Rev.}\ }\textbf {\bibinfo {volume} {244}},\ \bibinfo {pages}
  {115} (\bibinfo {year} {2003})}\BibitemShut {NoStop}%
\bibitem [{\citenamefont {Chen}\ \emph {et~al.}(2010)\citenamefont {Chen},
  \citenamefont {Fu},\ and\ \citenamefont {Franchini}}]{chen_2010}%
  \BibitemOpen
  \bibfield  {author} {\bibinfo {author} {\bibfnamefont {X.-Q.}\ \bibnamefont
  {Chen}}, \bibinfo {author} {\bibfnamefont {C.}~\bibnamefont {Fu}}, \ and\
  \bibinfo {author} {\bibfnamefont {C.}~\bibnamefont {Franchini}},\ }\href@noop
  {} {\bibfield  {journal} {\bibinfo  {journal} {J. Phys. Condens. Matter}\
  }\textbf {\bibinfo {volume} {22}},\ \bibinfo {pages} {292201} (\bibinfo
  {year} {2010})}\BibitemShut {NoStop}%
\bibitem [{\citenamefont {Benson}\ \emph {et~al.}(2013)\citenamefont {Benson},
  \citenamefont {Li}, \citenamefont {Luo}, \citenamefont {Ahuja}, \citenamefont
  {Svensson},\ and\ \citenamefont {H{\"a}ussermann}}]{benson_2013}%
  \BibitemOpen
  \bibfield  {author} {\bibinfo {author} {\bibfnamefont {D.}~\bibnamefont
  {Benson}}, \bibinfo {author} {\bibfnamefont {Y.-L.}\ \bibnamefont {Li}},
  \bibinfo {author} {\bibfnamefont {W.}~\bibnamefont {Luo}}, \bibinfo {author}
  {\bibfnamefont {R.}~\bibnamefont {Ahuja}}, \bibinfo {author} {\bibfnamefont
  {G.}~\bibnamefont {Svensson}}, \ and\ \bibinfo {author} {\bibfnamefont
  {U.}~\bibnamefont {H{\"a}ussermann}},\ }\href@noop {} {\bibfield  {journal}
  {\bibinfo  {journal} {Inorg. Chem.}\ }\textbf {\bibinfo {volume} {52}},\
  \bibinfo {pages} {6402} (\bibinfo {year} {2013})}\BibitemShut {NoStop}%
\bibitem [{\citenamefont {Li}\ \emph {et~al.}(2013)\citenamefont {Li},
  \citenamefont {Luo}, \citenamefont {Zeng}, \citenamefont {Lin}, \citenamefont
  {Mao},\ and\ \citenamefont {Ahuja}}]{li_2013}%
  \BibitemOpen
  \bibfield  {author} {\bibinfo {author} {\bibfnamefont {Y.-L.}\ \bibnamefont
  {Li}}, \bibinfo {author} {\bibfnamefont {W.}~\bibnamefont {Luo}}, \bibinfo
  {author} {\bibfnamefont {Z.}~\bibnamefont {Zeng}}, \bibinfo {author}
  {\bibfnamefont {H.-Q.}\ \bibnamefont {Lin}}, \bibinfo {author} {\bibfnamefont
  {H.-k.}\ \bibnamefont {Mao}}, \ and\ \bibinfo {author} {\bibfnamefont
  {R.}~\bibnamefont {Ahuja}},\ }\href@noop {} {\bibfield  {journal} {\bibinfo
  {journal} {Proc. Natl. Acad. Sci.}\ }\textbf {\bibinfo {volume} {110}},\
  \bibinfo {pages} {9289} (\bibinfo {year} {2013})}\BibitemShut {NoStop}%
\bibitem [{\citenamefont {Wang}\ \emph {et~al.}(2014)\citenamefont {Wang},
  \citenamefont {Zhou}, \citenamefont {Hu}, \citenamefont {Oganov},
  \citenamefont {Zhong},\ and\ \citenamefont {Rao}}]{wang_2014}%
  \BibitemOpen
  \bibfield  {author} {\bibinfo {author} {\bibfnamefont {D.-H.}\ \bibnamefont
  {Wang}}, \bibinfo {author} {\bibfnamefont {H.-Y.}\ \bibnamefont {Zhou}},
  \bibinfo {author} {\bibfnamefont {C.-H.}\ \bibnamefont {Hu}}, \bibinfo
  {author} {\bibfnamefont {A.~R.}\ \bibnamefont {Oganov}}, \bibinfo {author}
  {\bibfnamefont {Y.}~\bibnamefont {Zhong}}, \ and\ \bibinfo {author}
  {\bibfnamefont {G.-H.}\ \bibnamefont {Rao}},\ }\href@noop {} {\bibfield
  {journal} {\bibinfo  {journal} {Phys. Chem. Chem. Phys.}\ }\textbf {\bibinfo
  {volume} {16}},\ \bibinfo {pages} {20780} (\bibinfo {year}
  {2014})}\BibitemShut {NoStop}%
\bibitem [{\citenamefont {Li}\ \emph {et~al.}(2014)\citenamefont {Li},
  \citenamefont {Ahuja},\ and\ \citenamefont {Lin}}]{li_2014}%
  \BibitemOpen
  \bibfield  {author} {\bibinfo {author} {\bibfnamefont {Y.-L.}\ \bibnamefont
  {Li}}, \bibinfo {author} {\bibfnamefont {R.}~\bibnamefont {Ahuja}}, \ and\
  \bibinfo {author} {\bibfnamefont {H.-Q.}\ \bibnamefont {Lin}},\ }\href@noop
  {} {\bibfield  {journal} {\bibinfo  {journal} {Chin. Sci. Bull.}\ }\textbf
  {\bibinfo {volume} {59}},\ \bibinfo {pages} {5269} (\bibinfo {year}
  {2014})}\BibitemShut {NoStop}%
\bibitem [{\citenamefont {Li}\ \emph {et~al.}(2015)\citenamefont {Li},
  \citenamefont {Wang}, \citenamefont {Oganov}, \citenamefont {Gou},
  \citenamefont {Smith},\ and\ \citenamefont {Strobel}}]{li_2015}%
  \BibitemOpen
  \bibfield  {author} {\bibinfo {author} {\bibfnamefont {Y.-L.}\ \bibnamefont
  {Li}}, \bibinfo {author} {\bibfnamefont {S.-N.}\ \bibnamefont {Wang}},
  \bibinfo {author} {\bibfnamefont {A.~R.}\ \bibnamefont {Oganov}}, \bibinfo
  {author} {\bibfnamefont {H.}~\bibnamefont {Gou}}, \bibinfo {author}
  {\bibfnamefont {J.~S.}\ \bibnamefont {Smith}}, \ and\ \bibinfo {author}
  {\bibfnamefont {T.~A.}\ \bibnamefont {Strobel}},\ }\href@noop {} {\bibfield
  {journal} {\bibinfo  {journal} {Nat. Commun.}\ }\textbf {\bibinfo {volume}
  {6}},\ \bibinfo {pages} {6974} (\bibinfo {year} {2015})}\BibitemShut
  {NoStop}%
\bibitem [{\citenamefont {Efthimiopoulos}\ \emph
  {et~al.}(2012{\natexlab{a}})\citenamefont {Efthimiopoulos}, \citenamefont
  {Kunc}, \citenamefont {{G. V. Vazhenin}}, \citenamefont {Stavrou},
  \citenamefont {Syassen}, \citenamefont {Hanfland}, \citenamefont {{St.
  Liebig}},\ and\ \citenamefont {Ruschewitz}}]{ilias_2012}%
  \BibitemOpen
  \bibfield  {author} {\bibinfo {author} {\bibfnamefont {I.}~\bibnamefont
  {Efthimiopoulos}}, \bibinfo {author} {\bibfnamefont {K.}~\bibnamefont
  {Kunc}}, \bibinfo {author} {\bibnamefont {{G. V. Vazhenin}}}, \bibinfo
  {author} {\bibfnamefont {E.}~\bibnamefont {Stavrou}}, \bibinfo {author}
  {\bibfnamefont {K.}~\bibnamefont {Syassen}}, \bibinfo {author} {\bibfnamefont
  {M.}~\bibnamefont {Hanfland}}, \bibinfo {author} {\bibnamefont {{St.
  Liebig}}}, \ and\ \bibinfo {author} {\bibfnamefont {U.}~\bibnamefont
  {Ruschewitz}},\ }\href@noop {} {\bibfield  {journal} {\bibinfo  {journal}
  {Phys. Rev. B}\ }\textbf {\bibinfo {volume} {85}},\ \bibinfo {pages} {054105}
  (\bibinfo {year} {2012}{\natexlab{a}})}\BibitemShut {NoStop}%
\bibitem [{\citenamefont {Efthimiopoulos}\ \emph
  {et~al.}(2012{\natexlab{b}})\citenamefont {Efthimiopoulos}, \citenamefont
  {Vazhenin}, \citenamefont {Kunc}, \citenamefont {Stavrou}, \citenamefont
  {Syassen}, \citenamefont {Hanfland}, \citenamefont {Liebig},\ and\
  \citenamefont {Ruschewitz}}]{ilias_2012b}%
  \BibitemOpen
  \bibfield  {author} {\bibinfo {author} {\bibfnamefont {I.}~\bibnamefont
  {Efthimiopoulos}}, \bibinfo {author} {\bibfnamefont {G.~V.}\ \bibnamefont
  {Vazhenin}}, \bibinfo {author} {\bibfnamefont {K.}~\bibnamefont {Kunc}},
  \bibinfo {author} {\bibfnamefont {E.}~\bibnamefont {Stavrou}}, \bibinfo
  {author} {\bibfnamefont {K.}~\bibnamefont {Syassen}}, \bibinfo {author}
  {\bibfnamefont {M.}~\bibnamefont {Hanfland}}, \bibinfo {author}
  {\bibfnamefont {S.}~\bibnamefont {Liebig}}, \ and\ \bibinfo {author}
  {\bibfnamefont {U.}~\bibnamefont {Ruschewitz}},\ }in\ \href@noop {} {\emph
  {\bibinfo {booktitle} {Book of {Abstracts}}}}\ (\bibinfo {address} {50th
  EHPRG Meeting, Thessaloniki},\ \bibinfo {year} {2012})\ p.\ \bibinfo {pages}
  {224}\BibitemShut {NoStop}%
\bibitem [{\citenamefont {Efthimiopoulos}\ \emph {et~al.}(2010)\citenamefont
  {Efthimiopoulos}, \citenamefont {Vajenine}, \citenamefont {Stavrou},
  \citenamefont {Kunc}, \citenamefont {Syassen}, \citenamefont {Liebig},
  \citenamefont {Ruschewitz},\ and\ \citenamefont {Hanfland}}]{ilias_2010a}%
  \BibitemOpen
  \bibfield  {author} {\bibinfo {author} {\bibfnamefont {I.}~\bibnamefont
  {Efthimiopoulos}}, \bibinfo {author} {\bibfnamefont {G.}~\bibnamefont
  {Vajenine}}, \bibinfo {author} {\bibfnamefont {E.}~\bibnamefont {Stavrou}},
  \bibinfo {author} {\bibfnamefont {K.}~\bibnamefont {Kunc}}, \bibinfo {author}
  {\bibfnamefont {K.}~\bibnamefont {Syassen}}, \bibinfo {author} {\bibfnamefont
  {S.}~\bibnamefont {Liebig}}, \bibinfo {author} {\bibfnamefont
  {U.}~\bibnamefont {Ruschewitz}}, \ and\ \bibinfo {author} {\bibfnamefont
  {M.}~\bibnamefont {Hanfland}}\ }(\bibinfo {address} {European Crystallography
  Meeting (ECM26), Darmstadt},\ \bibinfo {year} {2010})\BibitemShut {NoStop}%
\bibitem [{\citenamefont {Nyl{\'e}n}\ \emph {et~al.}(2012)\citenamefont
  {Nyl{\'e}n}, \citenamefont {Konar}, \citenamefont {Lazor}, \citenamefont
  {Benson},\ and\ \citenamefont {H{\"a}ussermann}}]{nylen_2012}%
  \BibitemOpen
  \bibfield  {author} {\bibinfo {author} {\bibfnamefont {J.}~\bibnamefont
  {Nyl{\'e}n}}, \bibinfo {author} {\bibfnamefont {S.}~\bibnamefont {Konar}},
  \bibinfo {author} {\bibfnamefont {P.}~\bibnamefont {Lazor}}, \bibinfo
  {author} {\bibfnamefont {D.}~\bibnamefont {Benson}}, \ and\ \bibinfo {author}
  {\bibfnamefont {U.}~\bibnamefont {H{\"a}ussermann}},\ }\href@noop {}
  {\bibfield  {journal} {\bibinfo  {journal} {J. Chem. Phys.}\ }\textbf
  {\bibinfo {volume} {137}},\ \bibinfo {pages} {224507} (\bibinfo {year}
  {2012})}\BibitemShut {NoStop}%
\bibitem [{\citenamefont {Zhu}\ \emph {et~al.}(2012)\citenamefont {Zhu},
  \citenamefont {Oganov}, \citenamefont {Glass},\ and\ \citenamefont
  {Stokes}}]{zhu_2012}%
  \BibitemOpen
  \bibfield  {author} {\bibinfo {author} {\bibfnamefont {Q.}~\bibnamefont
  {Zhu}}, \bibinfo {author} {\bibfnamefont {A.~R.}\ \bibnamefont {Oganov}},
  \bibinfo {author} {\bibfnamefont {C.~W.}\ \bibnamefont {Glass}}, \ and\
  \bibinfo {author} {\bibfnamefont {H.~T.}\ \bibnamefont {Stokes}},\
  }\href@noop {} {\bibfield  {journal} {\bibinfo  {journal} {Acta Crystallogr.
  Sect. B}\ }\textbf {\bibinfo {volume} {68}},\ \bibinfo {pages} {215}
  (\bibinfo {year} {2012})}\BibitemShut {NoStop}%
\bibitem [{\citenamefont {Syassen}(2008)}]{syassen_2008}%
  \BibitemOpen
  \bibfield  {author} {\bibinfo {author} {\bibfnamefont {K.}~\bibnamefont
  {Syassen}},\ }\href@noop {} {\bibfield  {journal} {\bibinfo  {journal} {High
  Press. Res.}\ }\textbf {\bibinfo {volume} {28}},\ \bibinfo {pages} {75}
  (\bibinfo {year} {2008})}\BibitemShut {NoStop}%
\bibitem [{\citenamefont {Hammersley}\ \emph {et~al.}(1996)\citenamefont
  {Hammersley}, \citenamefont {Svensson}, \citenamefont {Hanfland},
  \citenamefont {Fitch},\ and\ \citenamefont {Hausermann}}]{hammersley_1996}%
  \BibitemOpen
  \bibfield  {author} {\bibinfo {author} {\bibfnamefont {A.~P.}\ \bibnamefont
  {Hammersley}}, \bibinfo {author} {\bibfnamefont {S.~O.}\ \bibnamefont
  {Svensson}}, \bibinfo {author} {\bibfnamefont {M.}~\bibnamefont {Hanfland}},
  \bibinfo {author} {\bibfnamefont {A.~N.}\ \bibnamefont {Fitch}}, \ and\
  \bibinfo {author} {\bibfnamefont {D.}~\bibnamefont {Hausermann}},\
  }\href@noop {} {\bibfield  {journal} {\bibinfo  {journal} {High Press. Res.}\
  }\textbf {\bibinfo {volume} {14}},\ \bibinfo {pages} {235} (\bibinfo {year}
  {1996})}\BibitemShut {NoStop}%
\bibitem [{\citenamefont {WinXPOW}(2010)}]{winxpow_2010}%
  \BibitemOpen
  \bibfield  {author} {\bibinfo {author} {\bibfnamefont {S.}~\bibnamefont
  {WinXPOW}},\ }\href@noop {} {\bibfield  {journal} {\bibinfo  {journal}
  {Darmstadt: Stoe \& Cie GmbH}\ } (\bibinfo {year} {2010})}\BibitemShut
  {NoStop}%
\bibitem [{\citenamefont {Boultif}\ and\ \citenamefont
  {Lou{\"e}r}(1991)}]{boultif_1991}%
  \BibitemOpen
  \bibfield  {author} {\bibinfo {author} {\bibfnamefont {A.}~\bibnamefont
  {Boultif}}\ and\ \bibinfo {author} {\bibfnamefont {D.}~\bibnamefont
  {Lou{\"e}r}},\ }\href@noop {} {\bibfield  {journal} {\bibinfo  {journal} {J.
  Appl. Crystallogr.}\ }\textbf {\bibinfo {volume} {24}},\ \bibinfo {pages}
  {987} (\bibinfo {year} {1991})}\BibitemShut {NoStop}%
\bibitem [{\citenamefont {Putz}\ \emph {et~al.}(1999)\citenamefont {Putz},
  \citenamefont {Sch{\"o}n},\ and\ \citenamefont {Jansen}}]{putz_1999}%
  \BibitemOpen
  \bibfield  {author} {\bibinfo {author} {\bibfnamefont {H.}~\bibnamefont
  {Putz}}, \bibinfo {author} {\bibfnamefont {J.}~\bibnamefont {Sch{\"o}n}}, \
  and\ \bibinfo {author} {\bibfnamefont {M.}~\bibnamefont {Jansen}},\
  }\href@noop {} {\bibfield  {journal} {\bibinfo  {journal} {J. Appl.
  Crystallogr.}\ }\textbf {\bibinfo {volume} {32}},\ \bibinfo {pages} {864}
  (\bibinfo {year} {1999})}\BibitemShut {NoStop}%
\bibitem [{\citenamefont {Larson}\ and\ \citenamefont
  {Von~Dreele}(2004)}]{larson_2004}%
  \BibitemOpen
  \bibfield  {author} {\bibinfo {author} {\bibfnamefont {A.~C.}\ \bibnamefont
  {Larson}}\ and\ \bibinfo {author} {\bibfnamefont {R.~B.}\ \bibnamefont
  {Von~Dreele}},\ }\href@noop {} {\emph {\bibinfo {title} {General {Structure}
  {Analysis} system ({GSAS})}}},\ \bibinfo {type} {Tech. Rep.}\ (\bibinfo
  {institution} {Los Alamos National Laboratory Report LAUR 86-748},\ \bibinfo
  {year} {2004})\BibitemShut {NoStop}%
\bibitem [{\citenamefont {Oganov}\ and\ \citenamefont
  {Glass}(2006)}]{oganov_2006}%
  \BibitemOpen
  \bibfield  {author} {\bibinfo {author} {\bibfnamefont {A.~R.}\ \bibnamefont
  {Oganov}}\ and\ \bibinfo {author} {\bibfnamefont {C.~W.}\ \bibnamefont
  {Glass}},\ }\href@noop {} {\bibfield  {journal} {\bibinfo  {journal} {J.
  Chem. Phys.}\ }\textbf {\bibinfo {volume} {124}},\ \bibinfo {pages} {244704}
  (\bibinfo {year} {2006})}\BibitemShut {NoStop}%
\bibitem [{\citenamefont {Glass}\ \emph {et~al.}(2006)\citenamefont {Glass},
  \citenamefont {Oganov},\ and\ \citenamefont {Hansen}}]{glass_2006}%
  \BibitemOpen
  \bibfield  {author} {\bibinfo {author} {\bibfnamefont {C.~W.}\ \bibnamefont
  {Glass}}, \bibinfo {author} {\bibfnamefont {A.~R.}\ \bibnamefont {Oganov}}, \
  and\ \bibinfo {author} {\bibfnamefont {N.}~\bibnamefont {Hansen}},\
  }\href@noop {} {\bibfield  {journal} {\bibinfo  {journal} {Comput. Phys.
  Commun.}\ }\textbf {\bibinfo {volume} {175}},\ \bibinfo {pages} {713}
  (\bibinfo {year} {2006})}\BibitemShut {NoStop}%
\bibitem [{\citenamefont {Lyakhov}\ \emph {et~al.}(2013)\citenamefont
  {Lyakhov}, \citenamefont {Oganov}, \citenamefont {Stokes},\ and\
  \citenamefont {Zhu}}]{lyakhov_2013}%
  \BibitemOpen
  \bibfield  {author} {\bibinfo {author} {\bibfnamefont {A.~O.}\ \bibnamefont
  {Lyakhov}}, \bibinfo {author} {\bibfnamefont {A.~R.}\ \bibnamefont {Oganov}},
  \bibinfo {author} {\bibfnamefont {H.~T.}\ \bibnamefont {Stokes}}, \ and\
  \bibinfo {author} {\bibfnamefont {Q.}~\bibnamefont {Zhu}},\ }\href@noop {}
  {\bibfield  {journal} {\bibinfo  {journal} {Comput. Phys. Commun.}\ }\textbf
  {\bibinfo {volume} {184}},\ \bibinfo {pages} {1172} (\bibinfo {year}
  {2013})}\BibitemShut {NoStop}%
\bibitem [{\citenamefont {Hoft}\ \emph {et~al.}(2006)\citenamefont {Hoft},
  \citenamefont {Gale},\ and\ \citenamefont {Ford}}]{hoft_2006}%
  \BibitemOpen
  \bibfield  {author} {\bibinfo {author} {\bibfnamefont {R.~C.}\ \bibnamefont
  {Hoft}}, \bibinfo {author} {\bibfnamefont {J.~D.}\ \bibnamefont {Gale}}, \
  and\ \bibinfo {author} {\bibfnamefont {M.~J.}\ \bibnamefont {Ford}},\
  }\href@noop {} {\bibfield  {journal} {\bibinfo  {journal} {Mol. Simul.}\
  }\textbf {\bibinfo {volume} {32}},\ \bibinfo {pages} {595} (\bibinfo {year}
  {2006})}\BibitemShut {NoStop}%
\bibitem [{\citenamefont {Soler}\ \emph {et~al.}(2002)\citenamefont {Soler},
  \citenamefont {Artacho}, \citenamefont {Gale}, \citenamefont {Garc{\'\i}a},
  \citenamefont {Junquera}, \citenamefont {Ordej{\'o}n},\ and\ \citenamefont
  {S{\'a}nchez-Portal}}]{soler_2002}%
  \BibitemOpen
  \bibfield  {author} {\bibinfo {author} {\bibfnamefont {J.~M.}\ \bibnamefont
  {Soler}}, \bibinfo {author} {\bibfnamefont {E.}~\bibnamefont {Artacho}},
  \bibinfo {author} {\bibfnamefont {J.~D.}\ \bibnamefont {Gale}}, \bibinfo
  {author} {\bibfnamefont {A.}~\bibnamefont {Garc{\'\i}a}}, \bibinfo {author}
  {\bibfnamefont {J.}~\bibnamefont {Junquera}}, \bibinfo {author}
  {\bibfnamefont {P.}~\bibnamefont {Ordej{\'o}n}}, \ and\ \bibinfo {author}
  {\bibfnamefont {D.}~\bibnamefont {S{\'a}nchez-Portal}},\ }\href@noop {}
  {\bibfield  {journal} {\bibinfo  {journal} {J. Phys. Condens. Matter}\
  }\textbf {\bibinfo {volume} {14}},\ \bibinfo {pages} {2745} (\bibinfo {year}
  {2002})}\BibitemShut {NoStop}%
\bibitem [{\citenamefont {Perdew}\ \emph {et~al.}(1996)\citenamefont {Perdew},
  \citenamefont {Burke},\ and\ \citenamefont {Ernzerhof}}]{perdew_1996}%
  \BibitemOpen
  \bibfield  {author} {\bibinfo {author} {\bibfnamefont {J.~P.}\ \bibnamefont
  {Perdew}}, \bibinfo {author} {\bibfnamefont {K.}~\bibnamefont {Burke}}, \
  and\ \bibinfo {author} {\bibfnamefont {M.}~\bibnamefont {Ernzerhof}},\
  }\href@noop {} {\bibfield  {journal} {\bibinfo  {journal} {Phys. Rev. Lett.}\
  }\textbf {\bibinfo {volume} {77}},\ \bibinfo {pages} {3865} (\bibinfo {year}
  {1996})}\BibitemShut {NoStop}%
\bibitem [{\citenamefont {Troullier}\ and\ \citenamefont
  {Martins}(1991)}]{troullier_1991}%
  \BibitemOpen
  \bibfield  {author} {\bibinfo {author} {\bibfnamefont {N.}~\bibnamefont
  {Troullier}}\ and\ \bibinfo {author} {\bibfnamefont {J.~L.}\ \bibnamefont
  {Martins}},\ }\href@noop {} {\bibfield  {journal} {\bibinfo  {journal} {Phys.
  Rev. B}\ }\textbf {\bibinfo {volume} {43}},\ \bibinfo {pages} {1993}
  (\bibinfo {year} {1991})}\BibitemShut {NoStop}%
\bibitem [{\citenamefont {Kresse}\ and\ \citenamefont
  {Furthm{\"u}ller}(1996)}]{kresse_1996}%
  \BibitemOpen
  \bibfield  {author} {\bibinfo {author} {\bibfnamefont {G.}~\bibnamefont
  {Kresse}}\ and\ \bibinfo {author} {\bibfnamefont {J.}~\bibnamefont
  {Furthm{\"u}ller}},\ }\href@noop {} {\bibfield  {journal} {\bibinfo
  {journal} {Phys. Rev. B}\ }\textbf {\bibinfo {volume} {54}},\ \bibinfo
  {pages} {11169} (\bibinfo {year} {1996})}\BibitemShut {NoStop}%
\bibitem [{\citenamefont {Bl{\"o}chl}(1994)}]{blochl_1994a}%
  \BibitemOpen
  \bibfield  {author} {\bibinfo {author} {\bibfnamefont {P.~E.}\ \bibnamefont
  {Bl{\"o}chl}},\ }\href@noop {} {\bibfield  {journal} {\bibinfo  {journal}
  {Phys. Rev. B}\ }\textbf {\bibinfo {volume} {50}},\ \bibinfo {pages} {17953}
  (\bibinfo {year} {1994})}\BibitemShut {NoStop}%
\bibitem [{\citenamefont {Kresse}\ and\ \citenamefont
  {Joubert}(1999)}]{kresse_1999}%
  \BibitemOpen
  \bibfield  {author} {\bibinfo {author} {\bibfnamefont {G.}~\bibnamefont
  {Kresse}}\ and\ \bibinfo {author} {\bibfnamefont {D.}~\bibnamefont
  {Joubert}},\ }\href@noop {} {\bibfield  {journal} {\bibinfo  {journal} {Phys.
  Rev. B}\ }\textbf {\bibinfo {volume} {59}},\ \bibinfo {pages} {1758}
  (\bibinfo {year} {1999})}\BibitemShut {NoStop}%
\bibitem [{\citenamefont {Monkhorst}\ and\ \citenamefont
  {Pack}(1976)}]{monkhorst_1976}%
  \BibitemOpen
  \bibfield  {author} {\bibinfo {author} {\bibfnamefont {H.~J.}\ \bibnamefont
  {Monkhorst}}\ and\ \bibinfo {author} {\bibfnamefont {J.~D.}\ \bibnamefont
  {Pack}},\ }\href@noop {} {\bibfield  {journal} {\bibinfo  {journal} {Phys.
  Rev. B}\ }\textbf {\bibinfo {volume} {13}},\ \bibinfo {pages} {5188}
  (\bibinfo {year} {1976})}\BibitemShut {NoStop}%
\bibitem [{\citenamefont {Bl{\"o}chl}\ \emph {et~al.}(1994)\citenamefont
  {Bl{\"o}chl}, \citenamefont {Jepsen},\ and\ \citenamefont
  {Andersen}}]{blochl_1994b}%
  \BibitemOpen
  \bibfield  {author} {\bibinfo {author} {\bibfnamefont {P.~E.}\ \bibnamefont
  {Bl{\"o}chl}}, \bibinfo {author} {\bibfnamefont {O.}~\bibnamefont {Jepsen}},
  \ and\ \bibinfo {author} {\bibfnamefont {O.~K.}\ \bibnamefont {Andersen}},\
  }\href@noop {} {\bibfield  {journal} {\bibinfo  {journal} {Phys. Rev. B}\
  }\textbf {\bibinfo {volume} {49}},\ \bibinfo {pages} {16223} (\bibinfo {year}
  {1994})}\BibitemShut {NoStop}%
\bibitem [{\citenamefont {Ruschewitz}\ \emph {et~al.}(2001)\citenamefont
  {Ruschewitz}, \citenamefont {M{\"u}ller},\ and\ \citenamefont
  {Kockelmann}}]{ruschewitz_2001}%
  \BibitemOpen
  \bibfield  {author} {\bibinfo {author} {\bibfnamefont {U.}~\bibnamefont
  {Ruschewitz}}, \bibinfo {author} {\bibfnamefont {P.}~\bibnamefont
  {M{\"u}ller}}, \ and\ \bibinfo {author} {\bibfnamefont {W.}~\bibnamefont
  {Kockelmann}},\ }\href@noop {} {\bibfield  {journal} {\bibinfo  {journal} {Z.
  Anorg. Allg. Chem.}\ }\textbf {\bibinfo {volume} {627}},\ \bibinfo {pages}
  {513} (\bibinfo {year} {2001})}\BibitemShut {NoStop}%
\bibitem [{not()}]{notex}%
  \BibitemOpen
  \href@noop {} {}\bibinfo {note} {See Supplemental Material at [URL] for (1)
  DFT optimized structure parameters for \textit{Immm}-, \textit{Pnma}- and
  \textit{Cmcm}-Li$_2$C$_2$ for the pressure range 0-40 GPa, (2) details of the
  structural investigation and selected crystallographic parameters of
  Li$_2$C$_2$ at 18.7 GPa, (3) comparison of DFT optimized and experimental
  structure parameters (Rietveld fits) at 7.2, 18.1, 18.7 and 19.3 GPa, (4)
  experimental lattice parameters for \textit{Immm}- and
  \textit{Pnma}-Li$_2$C$_2$ (Le Bail fits), (5) dispersion relation and
  corresponding phonon density of states of \textit{Pnma}-Li$_2$C$_2$ at 20
  GPa, (6) enthalpy-pressure relations of \textit{Pnma}-Li$_2$C$_2$ and
  Li$_2$C$_2$ with a CrB-type polycarbide structure}\BibitemShut {NoStop}%
\bibitem [{\citenamefont {Grzechnik}\ \emph {et~al.}(2000)\citenamefont
  {Grzechnik}, \citenamefont {Vegas}, \citenamefont {Syassen}, \citenamefont
  {Loa}, \citenamefont {Hanfland},\ and\ \citenamefont
  {Jansen}}]{grzechnik_2000}%
  \BibitemOpen
  \bibfield  {author} {\bibinfo {author} {\bibfnamefont {A.}~\bibnamefont
  {Grzechnik}}, \bibinfo {author} {\bibfnamefont {A.}~\bibnamefont {Vegas}},
  \bibinfo {author} {\bibfnamefont {K.}~\bibnamefont {Syassen}}, \bibinfo
  {author} {\bibfnamefont {I.}~\bibnamefont {Loa}}, \bibinfo {author}
  {\bibfnamefont {M.}~\bibnamefont {Hanfland}}, \ and\ \bibinfo {author}
  {\bibfnamefont {M.}~\bibnamefont {Jansen}},\ }\href@noop {} {\bibfield
  {journal} {\bibinfo  {journal} {J. Solid State Chem.}\ }\textbf {\bibinfo
  {volume} {154}},\ \bibinfo {pages} {603} (\bibinfo {year}
  {2000})}\BibitemShut {NoStop}%
\bibitem [{\citenamefont {Birch}(1947)}]{birch_1947}%
  \BibitemOpen
  \bibfield  {author} {\bibinfo {author} {\bibfnamefont {F.}~\bibnamefont
  {Birch}},\ }\href@noop {} {\bibfield  {journal} {\bibinfo  {journal} {Phys.
  Rev.}\ }\textbf {\bibinfo {volume} {71}},\ \bibinfo {pages} {809} (\bibinfo
  {year} {1947})}\BibitemShut {NoStop}%
\bibitem [{\citenamefont {Liebig}\ \emph {et~al.}(2013)\citenamefont {Liebig},
  \citenamefont {Paulus}, \citenamefont {Sternemann},\ and\ \citenamefont
  {Ruschewitz}}]{liebig_2013}%
  \BibitemOpen
  \bibfield  {author} {\bibinfo {author} {\bibfnamefont {S.}~\bibnamefont
  {Liebig}}, \bibinfo {author} {\bibfnamefont {M.}~\bibnamefont {Paulus}},
  \bibinfo {author} {\bibfnamefont {C.}~\bibnamefont {Sternemann}}, \ and\
  \bibinfo {author} {\bibfnamefont {U.}~\bibnamefont {Ruschewitz}},\
  }\href@noop {} {\bibfield  {journal} {\bibinfo  {journal} {Z. Anorg. Allg.
  Chem.}\ }\textbf {\bibinfo {volume} {639}},\ \bibinfo {pages} {2804}
  (\bibinfo {year} {2013})}\BibitemShut {NoStop}%
\bibitem [{\citenamefont {Kunc}\ \emph {et~al.}(2005)\citenamefont {Kunc},
  \citenamefont {Loa}, \citenamefont {Grzechnik},\ and\ \citenamefont
  {Syassen}}]{kunc_2005}%
  \BibitemOpen
  \bibfield  {author} {\bibinfo {author} {\bibfnamefont {K.}~\bibnamefont
  {Kunc}}, \bibinfo {author} {\bibfnamefont {I.}~\bibnamefont {Loa}}, \bibinfo
  {author} {\bibfnamefont {A.}~\bibnamefont {Grzechnik}}, \ and\ \bibinfo
  {author} {\bibfnamefont {K.}~\bibnamefont {Syassen}},\ }\href@noop {}
  {\bibfield  {journal} {\bibinfo  {journal} {Phys. Status Solidi B}\ }\textbf
  {\bibinfo {volume} {242}},\ \bibinfo {pages} {1857} (\bibinfo {year}
  {2005})}\BibitemShut {NoStop}%
\bibitem [{\citenamefont {Vegas}\ \emph {et~al.}(2001)\citenamefont {Vegas},
  \citenamefont {Grzechnik}, \citenamefont {Syassen}, \citenamefont {Loa},
  \citenamefont {Hanfland},\ and\ \citenamefont {Jansen}}]{vegas_2001}%
  \BibitemOpen
  \bibfield  {author} {\bibinfo {author} {\bibfnamefont {A.}~\bibnamefont
  {Vegas}}, \bibinfo {author} {\bibfnamefont {A.}~\bibnamefont {Grzechnik}},
  \bibinfo {author} {\bibfnamefont {K.}~\bibnamefont {Syassen}}, \bibinfo
  {author} {\bibfnamefont {I.}~\bibnamefont {Loa}}, \bibinfo {author}
  {\bibfnamefont {M.}~\bibnamefont {Hanfland}}, \ and\ \bibinfo {author}
  {\bibfnamefont {M.}~\bibnamefont {Jansen}},\ }\href@noop {} {\bibfield
  {journal} {\bibinfo  {journal} {Acta Crystallogr. Sect. B}\ }\textbf
  {\bibinfo {volume} {57}},\ \bibinfo {pages} {151} (\bibinfo {year}
  {2001})}\BibitemShut {NoStop}%
\bibitem [{\citenamefont {Vegas}\ and\ \citenamefont
  {Jansen}(2002)}]{vegas_2002}%
  \BibitemOpen
  \bibfield  {author} {\bibinfo {author} {\bibfnamefont {A.}~\bibnamefont
  {Vegas}}\ and\ \bibinfo {author} {\bibfnamefont {M.}~\bibnamefont {Jansen}},\
  }\href@noop {} {\bibfield  {journal} {\bibinfo  {journal} {Acta Crystallogr.
  Sect. B}\ }\textbf {\bibinfo {volume} {58}},\ \bibinfo {pages} {38} (\bibinfo
  {year} {2002})}\BibitemShut {NoStop}%
\bibitem [{\citenamefont {Santamaria-Perez}\ \emph {et~al.}(2011)\citenamefont
  {Santamaria-Perez}, \citenamefont {Vegas}, \citenamefont {Muehle},\ and\
  \citenamefont {Jansen}}]{santamaria_2011}%
  \BibitemOpen
  \bibfield  {author} {\bibinfo {author} {\bibfnamefont {D.}~\bibnamefont
  {Santamaria-Perez}}, \bibinfo {author} {\bibfnamefont {A.}~\bibnamefont
  {Vegas}}, \bibinfo {author} {\bibfnamefont {C.}~\bibnamefont {Muehle}}, \
  and\ \bibinfo {author} {\bibfnamefont {M.}~\bibnamefont {Jansen}},\
  }\href@noop {} {\bibfield  {journal} {\bibinfo  {journal} {Acta Crystallogr.
  Sect. B}\ }\textbf {\bibinfo {volume} {67}},\ \bibinfo {pages} {109}
  (\bibinfo {year} {2011})}\BibitemShut {NoStop}%
\bibitem [{\citenamefont {Hyde}\ and\ \citenamefont
  {Andersson}(1989)}]{hyde_1989}%
  \BibitemOpen
  \bibfield  {author} {\bibinfo {author} {\bibfnamefont {B.~G.}\ \bibnamefont
  {Hyde}}\ and\ \bibinfo {author} {\bibfnamefont {S.}~\bibnamefont
  {Andersson}},\ }\href@noop {} {\emph {\bibinfo {title} {Inorganic Crystal
  Structures}}}\ (\bibinfo  {publisher} {Wiley},\ \bibinfo {year}
  {1989})\BibitemShut {NoStop}%
\bibitem [{\citenamefont {Ruiz}\ and\ \citenamefont
  {Alemany}(1995)}]{ruiz_1995}%
  \BibitemOpen
  \bibfield  {author} {\bibinfo {author} {\bibfnamefont {E.}~\bibnamefont
  {Ruiz}}\ and\ \bibinfo {author} {\bibfnamefont {P.}~\bibnamefont {Alemany}},\
  }\href@noop {} {\bibfield  {journal} {\bibinfo  {journal} {J. Phys. Chem.}\
  }\textbf {\bibinfo {volume} {99}},\ \bibinfo {pages} {3114} (\bibinfo {year}
  {1995})}\BibitemShut {NoStop}%
\bibitem [{\citenamefont {Long}\ \emph {et~al.}(1992)\citenamefont {Long},
  \citenamefont {Hoffmann},\ and\ \citenamefont {Meyer}}]{long_1992}%
  \BibitemOpen
  \bibfield  {author} {\bibinfo {author} {\bibfnamefont {J.~R.}\ \bibnamefont
  {Long}}, \bibinfo {author} {\bibfnamefont {R.}~\bibnamefont {Hoffmann}}, \
  and\ \bibinfo {author} {\bibfnamefont {H.~J.}\ \bibnamefont {Meyer}},\
  }\href@noop {} {\bibfield  {journal} {\bibinfo  {journal} {Inorg. Chem.}\
  }\textbf {\bibinfo {volume} {31}},\ \bibinfo {pages} {1734} (\bibinfo {year}
  {1992})}\BibitemShut {NoStop}%
\bibitem [{\citenamefont {Sch{\"o}n}\ \emph {et~al.}(2004)\citenamefont
  {Sch{\"o}n}, \citenamefont {{\v{C}}an{\v{c}}arevi{\'c}},\ and\ \citenamefont
  {Jansen}}]{schon_2004}%
  \BibitemOpen
  \bibfield  {author} {\bibinfo {author} {\bibfnamefont {J.}~\bibnamefont
  {Sch{\"o}n}}, \bibinfo {author} {\bibfnamefont {{\v{Z}}.}~\bibnamefont
  {{\v{C}}an{\v{c}}arevi{\'c}}}, \ and\ \bibinfo {author} {\bibfnamefont
  {M.}~\bibnamefont {Jansen}},\ }\href@noop {} {\bibfield  {journal} {\bibinfo
  {journal} {J. Chem. Phys.}\ }\textbf {\bibinfo {volume} {121}},\ \bibinfo
  {pages} {2289} (\bibinfo {year} {2004})}\BibitemShut {NoStop}%
\bibitem [{\citenamefont {Yamaoka}\ \emph {et~al.}(1980)\citenamefont
  {Yamaoka}, \citenamefont {Shimomura}, \citenamefont {Nakazawa},\ and\
  \citenamefont {Fukunaga}}]{yamaoka_1980}%
  \BibitemOpen
  \bibfield  {author} {\bibinfo {author} {\bibfnamefont {S.}~\bibnamefont
  {Yamaoka}}, \bibinfo {author} {\bibfnamefont {O.}~\bibnamefont {Shimomura}},
  \bibinfo {author} {\bibfnamefont {H.}~\bibnamefont {Nakazawa}}, \ and\
  \bibinfo {author} {\bibfnamefont {O.}~\bibnamefont {Fukunaga}},\ }\href@noop
  {} {\bibfield  {journal} {\bibinfo  {journal} {Solid State Commun.}\ }\textbf
  {\bibinfo {volume} {33}},\ \bibinfo {pages} {87} (\bibinfo {year}
  {1980})}\BibitemShut {NoStop}%
\bibitem [{\citenamefont {Lu}\ and\ \citenamefont {Zhao}(2014)}]{lu_2014}%
  \BibitemOpen
  \bibfield  {author} {\bibinfo {author} {\bibfnamefont {Y.-L.}\ \bibnamefont
  {Lu}}\ and\ \bibinfo {author} {\bibfnamefont {H.}~\bibnamefont {Zhao}},\
  }\href@noop {} {\bibfield  {journal} {\bibinfo  {journal} {Mod. Phys. Lett.
  B}\ }\textbf {\bibinfo {volume} {28}},\ \bibinfo {pages} {1450190} (\bibinfo
  {year} {2014})}\BibitemShut {NoStop}%
\bibitem [{\citenamefont {Kulkarni}\ \emph {et~al.}(2010)\citenamefont
  {Kulkarni}, \citenamefont {Doll}, \citenamefont {Sch{\"o}n},\ and\
  \citenamefont {Jansen}}]{kulkarni_2010}%
  \BibitemOpen
  \bibfield  {author} {\bibinfo {author} {\bibfnamefont {A.}~\bibnamefont
  {Kulkarni}}, \bibinfo {author} {\bibfnamefont {K.}~\bibnamefont {Doll}},
  \bibinfo {author} {\bibfnamefont {J.}~\bibnamefont {Sch{\"o}n}}, \ and\
  \bibinfo {author} {\bibfnamefont {M.}~\bibnamefont {Jansen}},\ }\href@noop {}
  {\bibfield  {journal} {\bibinfo  {journal} {J. Phys. Chem B}\ }\textbf
  {\bibinfo {volume} {114}},\ \bibinfo {pages} {15573} (\bibinfo {year}
  {2010})}\BibitemShut {NoStop}%
\bibitem [{\citenamefont {Syassen}(1985)}]{syassen_1985}%
  \BibitemOpen
  \bibfield  {author} {\bibinfo {author} {\bibfnamefont {K.}~\bibnamefont
  {Syassen}},\ }\href@noop {} {\bibfield  {journal} {\bibinfo  {journal} {Phys.
  Status Solidi A}\ }\textbf {\bibinfo {volume} {91}},\ \bibinfo {pages} {11}
  (\bibinfo {year} {1985})}\BibitemShut {NoStop}%
\bibitem [{\citenamefont {Luo}\ \emph {et~al.}(1994)\citenamefont {Luo},
  \citenamefont {Greene}, \citenamefont {Ghandehari}, \citenamefont {Li},\ and\
  \citenamefont {Ruoff}}]{luo_1994}%
  \BibitemOpen
  \bibfield  {author} {\bibinfo {author} {\bibfnamefont {H.}~\bibnamefont
  {Luo}}, \bibinfo {author} {\bibfnamefont {R.~G.}\ \bibnamefont {Greene}},
  \bibinfo {author} {\bibfnamefont {K.}~\bibnamefont {Ghandehari}}, \bibinfo
  {author} {\bibfnamefont {T.}~\bibnamefont {Li}}, \ and\ \bibinfo {author}
  {\bibfnamefont {A.~L.}\ \bibnamefont {Ruoff}},\ }\href@noop {} {\bibfield
  {journal} {\bibinfo  {journal} {Phys. Rev. B}\ }\textbf {\bibinfo {volume}
  {50}},\ \bibinfo {pages} {16232} (\bibinfo {year} {1994})}\BibitemShut
  {NoStop}%
\end{thebibliography}%



\clearpage

\end{document}